\documentclass[12pt,preprint]{aastex}
\pdfoutput=1                        
\usepackage{emulateapj5} 
\usepackage{onecolfloat} 
\usepackage{apjfonts}
\usepackage{subfigure}  
\usepackage{color}
\def\aap{A\&A} 
\def\apj{ApJ} 
 
\def\apjl{ApJL} 
\def\mnras{MNRAS} 
\def\araa{ARA\&A}

\def\apjs{ApJS} 

\def\gca{Geochimica et Cosmochimica Acta} 
\def\prd{Physical Review D}

\newcommand{\bdm}{\begin{displaymath}} 
\newcommand{\edm}{\end{displaymath}}
\newcommand{\beq}{\begin{equation}} 
\newcommand{\eeq}{\end{equation}} 
\newcommand{\beqnarr}{\begin{eqnarray}}
\newcommand{\eeqnarr}{\end{eqnarray}}
\newcommand{\bit}{\begin{itemize}} 
\newcommand{\eit}{\end{itemize}} 
\newcommand{\ben}{\begin{enumerate}} 
\newcommand{\een}{\end{enumerate}}
\newcommand{\bfi}{\begin{figure}[htb]} 
\newcommand{\bpfi}{\begin{figure}[p]}
\newcommand{\barr}{\begin{array}}
\newcommand{\earr}{\end{array}}
\newcommand{\bec}{\begin{center}}
\newcommand{\eec}{\end{center}}
\newcommand{\bs}{\begin{sideways}}
\newcommand{\es}{\end{sideways}}

\newcommand{\lta}{\sim}
\newcommand{\gta}{\gtrsim}

\newcommand{\msun}{\>{M_{\odot}}}

\newcommand{\dmonly}{DM }
\newcommand{\ovisc}{NR }
\newcommand{\agn}{AGN }
\newcommand{\csf}{CSF }

\newcommand{\xray}{X-ray}
\newcommand{\simu}{{{\sc sim}}}

\def \hinv      {{\, {h}^{-1}}}

\def\myputfigure#1#2#3#4#5
{\hskip0.03\textwidth\vskip#5pt
\makebox[0pt]{\hskip#2in
\includegraphics[width=#3\textwidth]{#1}}\vskip#4pt\hfill} 
\def\spose#1{\hbox to 0pt{#1\hss}}
\def\lta{\mathrel{\spose{\lower 3pt\hbox{$\mathchar"218$}}
     \raise 2.0pt\hbox{$\mathchar"13C$}}}
\def\gta{\mathrel{\spose{\lower 3pt\hbox{$\mathchar"218$}}
     \raise 2.0pt\hbox{$\mathchar"13E$}}}

\shorttitle{X-ray $c-M$  relation}
\shortauthors{Rasia et al.}
\begin{document}
\twocolumn[%
\title{On the discrepancy between Theoretical and X-ray
  Concentration--Mass Relations for Galaxy Clusters}
\author{
E.~Rasia\altaffilmark{1},
S.~Borgani\altaffilmark{2,3,4},
S.~Ettori\altaffilmark{5,6},
P.~Mazzotta\altaffilmark{7,8}, AND
M.~Meneghetti\altaffilmark{5,6}
}
\affil{$^1$  Department of Physics, University of Michigan, 450 Church Street, Ann Arbor, MI  48109,USA}
\affil{$^2$ Dipartimento di Fisica dell' Universit\`a di Trieste, Sezione di Astronomia, via Tiepolo 11, I-34131 Trieste, Italy}
\affil{$^3$ INAF, Osservatorio Astronomico di Trieste, via Tiepolo 11, I-34131, Trieste, Italy}
\affil{$^4$ INFN, Instituto Nazionale di Fisica Nucleare, Trieste, Italy}
\affil{$^5$ INAF, Osservatorio Astronomico di Bologna, via Ranzani 1, I-40127, Bologna, Italy}
\affil{$^6$ INFN, Sezione di Bologna, viale Berti Pichat 6/2, I-40127, Bologna, Italy}
\affil{$^7$ Dipartimento di Fisica, Universit\`a di Roma Tor Vergata, via della Ricerca Scientifica, I-00133, Roma, Italy}
\affil{$^8$ Harvard-Smithsonian Center for Astrophysics, 60 Garden Street, Cambridge, MA 02138, USA}
\date{Draft 1: August, 2011}

\begin{abstract}
In the past 15 years, the concentration--mass relation has been investigated diffusely in theoretical studies. On the other hand, only recently has this relation been derived from X-ray observations. When that happened, the results caused a certain level of concern: the X-ray normalizations and slopes were found significantly dissimilar from those predicted by theory. 

We analyzed 52 galaxy clusters and groups, simulated with different
descriptions of the physical processes that affect the baryonic
component, with the purpose of determining whether these discrepancies
are real or induced by biases in the computation of the concentration
parameter or in the determination of the selection function of the
cluster sample for which the analysis is carried out.  In particular,
we investigate how the simulated concentration--mass relation depends
(1) on the radial range used to derive the concentration, (2) on
the presence of baryons in the simulations, and on the effect of star
formation and feedback from supernovae and active galactic nuclei (AGNs). Finally, we evaluate
(3) how the results differ when adopting an X-ray approach for the
analysis and (4) how the selection function based on X-ray
luminosity can impact the results. All
effects studied go in the direction of alleviating the discrepancy
between observations and simulations, although with different significance:
while the choice of the radial range to fit the profiles and the
inclusion of the baryonic component play only a minor role, the X-ray
approach to reconstruct the mass profiles and the selection of
  the cluster sample have a strong impact on the resulting
concentration--mass relation.

Extending the fit to the most central regions or reducing the fitting
radius from the virial boundary to the typical X-ray external radius
causes an increase of the normalization  in radiative simulations
  by 5\%-10\%. In the second case,  we measure a slope that is up
  to twice steeper than that derived by using the typical theoretical
  radial
  range. 
Radiative simulations including only supernova feedback produce 30\% 
higher concentrations than the dark matter case. Such a difference is
largely reduced when including the effect of AGN feedback. The
concentration--mass relation derived from the X-ray synthetic catalog
is significantly steeper due to the combination of several different
effects, such as environment, dynamical state and dynamical history of
the clusters, bias in mass and temperature measurements, and their
dependence on the radius and on the mass of the system.  Finally,
selecting clusters according to their X-ray luminosity produces a net
increase in both normalization and slope of the relation, since at fixed mass, the most
luminous clusters are also the most concentrated.
\end{abstract}
\begin{keywords}
{cosmology: theory -- galaxies: clusters: general --galaxies: clusters: intracluster medium -- methods: numerical -- X-ray: galaxies: clusters}
\end{keywords} 
]

\section{Introduction}\label{sec:intro} 

The dark sector (dark energy and dark matter (DM)) exceeds by about 25 times the baryonic content of our universe (e.g. \citealt{voit05,borgani_kravtsov,allen.etal.12} for recent reviews on cosmology with focus on galaxy clusters). As a consequence, at first approximation, we may assume that simulations with only dark components properly describe the salient halo properties such as the mass distribution, the tridimensional halo shape, the density profile, and the concentration \citep[see][ for a galaxy-cluster-formation review]{kravtsov&borgani12}. However, an increasing number of theoretical papers have begun to contest this simplistic view, showing that baryons (in the form of both hot gas and stars) can significantly influence all these topics \citep[e.g.][ among the most recent publications on the various aspects]{rudd.etal.08,stanek.etal.09,duffy.etal.10,lau.etal.11,gnedin.etal.11,vandaalen.etal.11,cui.etal.11,zhu.etal.12,maccio.etal.12,governato.etal.12,balaguera.etal.12}. A certain disagreement between some observational results and theoretical predictions based on DM-only simulations support this discomfort.
In this respect, a recent debate about the X-ray observed and simulated concentration--mass ($c-M$) relation warmed up \citep[][and references therein]{fedeli12}. 
Indeed, while earlier works based on 10-12 clusters observed by {\it Chandra} \citep{vikh.etal.06} and {\it XMM-Newton} \citep{pratt&arnaud, pointecouteau.etal.05} showed consistency with theoretical works \citep{dolag.etal.04}, lately, with the enlargement of the observational collections, more differences have arisen. The $c-M$ relation from the larger X-ray samples of \citet[][ hereafter B07]{buote.etal.07}, \citet[][ hereafter SA07]{schmidt_allen}, and \citet[][ hereafter E10]{ettori.etal.10} is significantly steeper \citep{fedeli12} than the relation derived from the Millennium simulation \citep[][ with results similar to the aforementioned work by \citealt{dolag.etal.04}]{gao.etal.08}. An exception to this mismatch is the result by \cite{host&hansen}, who, however, analyzed only 11 systems.

All these works consistently adopt the Navarro--Frenk--White (NFW) analytic model to fit the density profile and derive the concentration.  
\citet{nfw96} showed the existence of a universal profile that well describes the halo density and mass distribution for a large range of halo mass and cosmologies \citep{nfw}. Their fitting formulae are characterized by two parameters: the normalization, $\rho_s$, and the scale radius, $r_s$. The expressions for the profiles of the NFW density and the NFW mass,  as a function of $x= r/r_s$, are respectively 
\beqnarr
\rho(x)=\frac{\rho_s}{x(1+x)^2},\ \ \ \ \ \ \ \ \ \ \ \ \ \ \ \ \ \ \\
M(<x)= 4 \pi \rho_s r_s^3 f_x, \ \ \ f_x=\ln(1+x)-x/(1+x).
\label{eq:massnfw}
\eeqnarr

\ 

The  logarithmic slope of the density profile measured at the scale radius is equal to $-2$, transitioning from the central asymptotic value of $-1$ to the external value of $-3$. The normalization factor is connected to the characteristic over-density, $\Delta$, and to the critical density at the halo redshift $z$, $\rho_{\rm cr}$,  through
 \beq
 \label{eq:rhos}
 \rho_s=\frac{\Delta \rho_{\rm cr}}{3}\frac{c^3}{\ln(1+c)-c/(1+c)}
 \eeq
where the critical density can be expressed as $\rho_{\rm cr} = 3 H^2 (z) / (8  G)$, with $H(z)$ the Hubble parameter at redshift $z$ and $G$ Newton's gravitational constant. Finally, $c$ is the concentration of the halo and is defined as the number of times the scale radius is contained within a fixed over-density radius: $c=R_{\Delta}/{r_s}$. In the following, we will always define the concentration in relation to the over-density $\Delta = 200.$\footnote{$R_{200}$ is, therefore, the radius of the sphere enclosing a mean density 200 times the critical density. Conversions of the NFW mass defined for different values of $\Delta$ are provided in the Appendix of \cite{hu&kravtsov} and Appendix of \cite{ettori.etal.10}.} 

The origins of the density shape mostly reside on the density and triaxiality of the original peak and  on the continuous contraction of the innermost material due to subsequent accretion during the collapse \citep{dalal.etal.10}. This last point is also a suitable explanation for the increase of the concentration parameter with the decrease of mass  \citep[e.g., NFW,][]{eke.etal.01,bullock.etal.01,dolag.etal.04,zhao.etal.09}. The physical and cosmological justification is linked to the hierarchical scenario of structure formation: small halos form at earlier time when the universe was denser than today and, over time, the sedimentation of material at the center happens in an already-established high-density peak.

Measures of a negative slope from observations are, therefore,
expected. The surprise is related to the values: the X-ray slope is up
to five times steeper than what is found in DM cosmological
simulations.  The question we want to answer in this paper is {\it
    whether the differences between theory and observations are real}
  (i.e., some fundamental physical ingredients are missing in
  theoretical models) {\it or artificial}  
(i.e., the comparison is
  performed without properly accounting for the presence of biases
  induced by the specific method to select the sample or to measure
  the concentration).  Indeed, it is unquestionable
that the basic procedural assumptions of the two analyses differ in
several aspects. The {\it simulated concentration--mass relation} is
mostly derived by fitting the NFW profile to the three-dimensional
density distribution extracted considering the influence of all the
particles without exclusion of sub-structures or sub-clumps, and from
the really central region to the outskirts of the halo, usually up to
the virial region. The sample selection is typically
volume-limited with a well-defined cut in mass.  Finally, as said
above, many theoretical works are based on large cosmological boxes of
DM only \cite[e.g.,][ among the most recent ones]{bhattacharya.etal.11,
  kwan.etal.12}.  On the contrary, the {\it X-ray concentration--mass
  relation} is derived from projected data that have a limited radial
range because of the signal-to-noise ratio and the field of view. Few
clusters have been observed to $R_{500}$ with enough photon statistics
and even fewer at $R_{200}$. For example, E10 (one of the samples we
are comparing to) reaches on average 40\% of $R_{200}$ (see
Section~\ref{sec:cmrr}). The observational selection function is not
done in mass. In the best case, it depends on the X-ray luminosity. In
the worst (and the most common) situation, the clusters are chosen
among the ones available in the archives without a well-defined
selection function. As a final note, the real universe has baryons.

The possibility that baryons might influence the structure-formation process and, in particular, the DM distribution was introduced almost three decades ago with the development of the analytic model of  `adiabatic contraction' \citep {barnes.etal.84,blumenthal.etal.86,ryden&gunn}. More recently, cosmological simulations push further the improvement of the model  \citep[e.g.][]{gnedin.etal.04,gnedin.etal.11,gustafsson.etal.06, abadi.etal.10,tissera.etal.10,zhemp.etal.12,giocoli.etal.12}. The underlying idea is that, at the center, the DM feels the growth of the total potential well due to the extra baryonic material accreted as a consequence of cooling and condensation. However, even if a specific great effort has been devoted to studying the concentration parameter in non-radiative simulations \citep[e.g.][]{rasia.etal.04, pedrosa.etal.09,tissera.etal.10} only recently have these investigations been extended to include various models of the intra cluster medium  \citep[ICM;][]{ rudd.etal.08,duffy.etal.10,meneghetti.etal.10,maccio.etal.12,governato.etal.12, martinizzi.etal.12,ragone.etal.12} and also different dark-energy models \citep{grossi&springel,baldi12,deboni.etal.12}.
 As for the analysis of hydrodynamical simulations, particular
  attention has been devoted so far to galaxies and galaxy groups to understand whether 
  the inclusion of baryons can explain the observational
  presence of cores in the density profiles \citep[e.g.][]{ogiya.etal.11,
    maccio.etal.12,martinizzi.etal.12,teyssier.etal.12}. On the
  contrary, in this work, we will focus on {\it massive clusters} that
  cover the same mass range sampled by the X-ray observations
  analyzed in SA07 and
  E10.
  
This paper is organized as follows: after presenting the simulations in Section 2, we discuss the results derived from the intrinsic analysis of the simulated clusters in Section 3. The observational approach applied to synthetic X-ray images will be presented in Section 4, while in Section 5 we will focus on the selection function. Our conclusions are outlined in Section 6.

\section{Simulations} \label{sec:sim} 

In this section, we briefly describe the four sets of simulated
clusters that are used in our intrinsic analysis (Section~3). These four sets are obtained
starting from the same initial conditions and using four different
physical descriptions of the processes determining the evolution of the
baryonic component. More details on the generation of the initial
conditions are provided by \cite{bonafede.etal.11}, while we refer to
\cite{killedar.etal.12} for a description of the different physical
models implemented, and to \citet[][ hereafter R12]{rasia.etal.12} for the
characterization of the synthetic X-ray catalog.

\subsection{The Sets of Simulated Clusters}
Our simulated clusters have been chosen within 29 Lagrangian
regions identified within a $1\, {h}^{-3}$Gpc$^3$ low-resolution $N$-body
cosmological simulation.\footnote{${h}=0.72$ throughout the paper.} The cosmological model assumed is a flat
$\Lambda$CDM model, with $\Omega_m=0.24$ for the matter density
parameter, $\Omega_{\rm bar}=0.04$ for the contribution of baryons,
$H_0=72\,{\rm km\,s^{-1}Mpc^{-1}}$ for the present-day Hubble
constant, $n_s=0.96$ for the primordial spectral index and
$\sigma_8=0.8$ for the normalization of the power spectrum. 
Resolution is increased within the selected regions by enlarging the
number of particles and correspondingly adding higher-frequency modes
from the power spectrum of the same cosmological model, by using 
the Zoomed Initial Condition  technique
\citep{tormen.etal.97}.  The runs were carried out using the GADGET-3
code, a newer and more efficient version of the former GADGET-2 code
\citep{sp05.1}. In all simulations a Plummer-equivalent softening length
for the computation of the gravitational force in the high-resolution
region was fixed to $\epsilon=5\,h^{-1}$kpc in physical units for the most recent
redshifts ($z<2$), while it was kept fixed in comoving units at earlier
epochs. As for the computation of hydrodynamic forces, we assume the
smoothed particle hydrodynamics smoothing length to reach a minimum allowed value of
$0.5\epsilon$. 

 Besides a set of dDM-only simulations and a set of
non-radiative (NR) hydrodynamical simulations, we also carried
out two sets of radiative simulations (\csf and {{\sc agn}}).  In these simulations radiative
cooling rates are computed by following the procedure presented by
\cite{wiersma.etal.09}, including the effect of cosmic microwave background radiation and of
UV/X-ray background radiation from quasars and galaxies
\citep{haardt&madau01}. The contributions to cooling from 11
elements (H, He, C, N, O, Ne, Mg, Si, S, Ca, Fe) have been
pre-computed using {\sc CLOUDY}
\citep{ferland.etal.98} for an optically thin gas in
(photo-)ionization equilibrium.

The \csf set of simulations includes star formation and the
effect of feedback triggered by supernova (SN) explosions. As for the star formation model, gas particles above a
given threshold density are treated as multiphase, so as to provide a
sub-resolution description of the interstellar medium, according to
the model originally described by \cite{springel&hernquist03}.  SNe II,
SNe Ia, and low- and intermediate-mass stars contribute to the production
of metals according to the model described by \cite{tornatore.etal.07}:
assuming a Chabrier initial mass function \citep{chabrier03} stars produce metals over
the time scale determined by the mass-dependent life times of
\cite{padovani&matteucci93}.  The effect of kinetic feedback triggered
by SNe II is included according to the prescription by
\cite{springel&hernquist03}.  In the \csf simulation set we
assume ${v}_{\rm{w}} = 500$ km s$^{-1}$ for the velocity of the outflows,
with a mass-upload rate that is two times the value of the local star
formation rate.

The \agn set of simulations is carried out by including the same
physical processes as in the \csf case, with a lower wind
velocity of ${v}_{\rm{w}} = 350$ km s$^{-1}$, but also including the effect
of AGN feedback. In this model, largely
based on the original implementation of black-hole (BH) feedback by
\cite{springel.etal.05}, energy results from gas accretion onto
supermassive BHs.  BHs are included as sink particles,
which grow in mass by gas accretion and merging with other BHs. Gas
accretion proceeds at a Bondi rate, and is limited by the Eddington
rate. Once the accretion rate is computed for each BH particle, a
stochastic criterion is used to select the surrounding gas particles
to be accreted. Unlike in \citet{springel.etal.05}, in which a selected gas particle
contributes to accretion with all its mass, we included the
possibility for a gas particle to accrete only with a slice of its
mass, which corresponds to 1/4 of its original mass, thus providing a
more continuous description of the accretion process \citep[see
also][]{fabjan.etal.10}.  BH particles are initially seeded with a mass
of $5 \times 10^{8} {h}^{-1}M_\odot$. Seeding of
BH particles takes place at the minimum of the potential of halos when
they first reach a minimum friend-of-friend mass of $2.5\times
10^{13} {h}^{-1}M_\odot$ (using a linking length of 0.16 in units
of the mean inter-particle separation), with the further condition that
such halos should contain a minimum mass fraction in stars of
0.02. This condition guarantees that substantial star formation took
place in such halos and that seeding does not take place in structures
linked together by the friend-of-friend algorithm, not corresponding to a real
halo. A radiation efficiency parameter of $\epsilon_r= 0.1$ determines
the amount of radiated energy extracted from gas accretion, in units
of the rest-mass energy of the accreted gas, with the BH mass being
correspondingly decreased by this amount. A fraction $\epsilon_f$ of
this radiated energy is thermally coupled to the surrounding gas.  We
use $\epsilon_f = 0.1$ for this feedback efficiency, which increases
to $\epsilon_f = 0.4$ when accretion enters in the quiescent ``radio''
mode and takes place at a rate smaller than 1/100 of the
Eddington limit \citep[e.g.,][]{sijacki.etal.07, fabjan.etal.10}.

\subsection{Simulated Catalog for the Intrinsics Analysis}
\label{sec:catalogue}

The intrinsic analysis described in Section~3 is based on a set of 52
objects.  Only the 24 largest clusters constitute a complete sample 
being the most massive halos identified within the parent cosmological
box of 1 $h^{-1}$ Gpc. These were specifically selected for the
purpose of 
comparing simulations of galaxy clusters with X-ray observations that
are available for a fairly large number of massive, X-ray luminous objects.
In order to extend the mass range over which to measure the
$c-M$ relation, we include in our analysis also lower-mass systems, identified within a variety of different environments.
The final sample includes the central clusters of each of 29 Lagrangian
regions (24 centered on as many of the most massive clusters, plus 5
regions centered on lower-mass isolated systems), three halos selected
from rich environments surrounding the massive clusters (regions
containing more than 20 halos with virial mass larger than $5 \times
10^{13} h^{-1} M_{\odot}$), six halos from a poor environment
(Lagrangian regions containing less than eight halos per region with
mass larger than the same limit), and 14 halos lying in a medium-populated environment.  The sample-covered mass interval is similar to
that usually considered in observational works, with $M_{200}$ ranging
from ~$7.5\times 10^{13} \hinv \msun$ to ~$2.5\times 10^{15} \hinv
\msun$. All clusters have been analyzed at redshift 0 (for
comparison with other works present in literature) and 0.25 (for
comparison with the X-ray sample of R12). Throughout the paper, we
will focus on the $z=0$ objects reporting our $z=0.25$ results in the
Appendix.

\subsection{Synthetic Catalog}
\label{sec:catalogue}

The  synthetic X-ray catalog (R12) employed in Section~4  contains 60 event files related to three orthogonal projections of 20 massive  objects as part of the \csf sample at $z=0.25.$\footnote{Notice that X-ray synthetic catalog has a more restricted mass range with respect to the sample of the intrinsic analysis.}  The  catalog has been created using the {\it X-Ray MAp Simulator} \cite[X-MAS;][]{gardini.etal.04,rasia.etal.08} after the removal of the over-cooled particles identified in the density ($\rho[{\rm gr} /{\rm cm}^(-3)]$) $-$ temperature ($T$[keV]) plane that satisfy the condition $T< 3\times 10^6 \rho^{0.25}$ . This cut reduces the presence of small peaked clumps in the X-ray images without affecting the overall distributions of density, temperature, and X-ray luminosity (see Appendix of R12). The redistribution matrix function and ancillary response function adopted  are those of {\it Chandra} ACIS-S3. The field of view is 16 arcmin, equivalent to $\sim$ -2.5 $\hinv$ Mpc for our cosmology and redshift, and the exposure time is 100 ks. The metallicity is kept constant to 0.3 solar with respect to the tables of \cite{anders_grevesse} with a small correction on the helium abundances to be consistent with the simulated hydrogen mass fraction. The galactic absorption, described by a {\it WABS} model, is fixed to $N_{\rm H}=5 \times 10^{-20}$ cm$^{-2}$. Along the line of sight, we consider the information of all the particles located up to 5 $\hinv$ Mpc (in both directions) from the center of the object. The final event files avoid the inclusion of the background since  this component does not influence the mass profile derivation \citep{rasia.etal.06}.

\section{The intrinsic {\it c-}M relation} \label{sec:cmi} 

\subsection{Density Profiles and Fitting Procedure}
 \label{sec:profiles}
To perform the tri-dimensional analysis of our simulated sample, we proceed by extracting the spherical mass profiles in 50 bins logarithmically spaced from 10 ${h}^{-1}$ kpc to 5 ${h}^{-1}$ Mpc and centered on the minimum of the potential well. 
For each cluster, all the bins external to $\approx$ 2\% of $R_{200}$
have more than 1000 particles. This number represents the
threshold for numerical convergency of the inner slope
\citep{power.etal.03}.  
 Following an observational approach, we fit our simulated profile
  with the NFW mass expression (Equation (2)) assuming a 10\% error on
  the mass, a value consistent with the typical errors in observations
  (B07; E10) and previous analyses of synthetic X-ray catalogs
  \citep[][ R12]{meneghetti.etal.10}. Using the values of $\rho_s$ and
  $r_s$ obtained from the fit of the density profiles , we compute
  the NFW masses 
  at critical overdensity $200$.  We then compute the concentration using
  Equation (\ref{eq:rhos}). The 1 $\sigma$ statical errors on the NFW mass
  and concentration are derived through the propagation of the errors
  on $\rho_s$ and $r_s$ applied to Equations (2) and (3), respectively. The
  $c-M$ relation is, then, calculated using the {\sc
    linmix\_err} routine \citep{kelly.etal.07} in IDL to linearly fit
  the following expression: 

\begin{equation}
\log_{10} c = \log_{10} c_0 + \alpha \times \log_{10} \frac{M_{\rm NFW}}{5\times 10^{14}\hinv \msun} + \epsilon_{\log}
\label{eq:cm}
\end{equation}

\noindent  The routine includes measurement-error information in both variables, utilizes a Bayesian approach, and returns a posteriori distributions of normalization ($\log_{10} c_0$), slope ($\alpha$), and the variance of the intrinsic scatter ($\epsilon_{\log}$) assumed to be log-normal \citep{dolag.etal.04,deboni.etal.12} with mean equal to zero. 
For the reduced size of our sample we used the flag enabling the Metropolitan--Hasting algorithm. From the distributions,  the median values and their error, $\sigma$ defined as the half-distance from the two values containing 68.27\% of the distribution,\footnote{For a symmetric distribution this would be equivalent to the standard deviation.} are extracted. Subsequently, the normalization and its statistical error are converted from logarithm to linear scale: $c_0=10^{log_{10}c_0}$ and $\sigma_{c_0} = \ln(10)  \times c_0 \times \sigma_{log_{10} c_0}$.  Finally, the statistical scatter is computed as
\begin{equation}
\sigma_{\rm stat}=\sqrt{\frac{\sum_i [\sigma_{\log_{10}(c_i)}]^2 }{N}}=\sqrt{\frac{\sum_i [\sigma_{(c_i)}/c_i/\ln(10)]^2 }{N}},
\end{equation}
 where $c_i$ is the concentration of the $i$ cluster, $\sigma_{(c,i)}$ its statistical error, and $N$ the number of clusters considered.

The process is performed multiple times on the same object changing in each circumstance the radial range assumed (Section~3.3). 
For the hydrodynamical simulations (\ovisc, \csf, and {\sc agn}), we analyze both the total mass profile and the DM-only profile. 

\subsection{Redefinition of the Sample}
When studying the influence of the radial range (Section~3.3) and the baryonic physics (Section~\ref{sec:bar}), we consider {\it only} the profiles that show a good consistency with the NFW description. In this way we exclude both objects that present one or more large substructures that produce a secondary peak in the density profile and halos that do not have a coherent concentration parameter associated. This a priori selection is not usually embraced in theoretical works, however, it is often implicitly adopted in observational samples, especially if they are focused on regular systems. 
The careful choice of the simulated clusters that need to be discarded is done on the basis of the NFW residuals:
\begin{equation}
\sigma_{\rm res}=\sqrt{\frac{\sum_i^{N_{\rm bin}} [\log_{10}(M_i)-\log_{10}(M_{\rm NFW,i})]^2 }{N_{\rm bin}}}.
\label{eq:residual}
\end{equation}
\begin{figure}[h]
\centering
\includegraphics[width=0.48\textwidth]{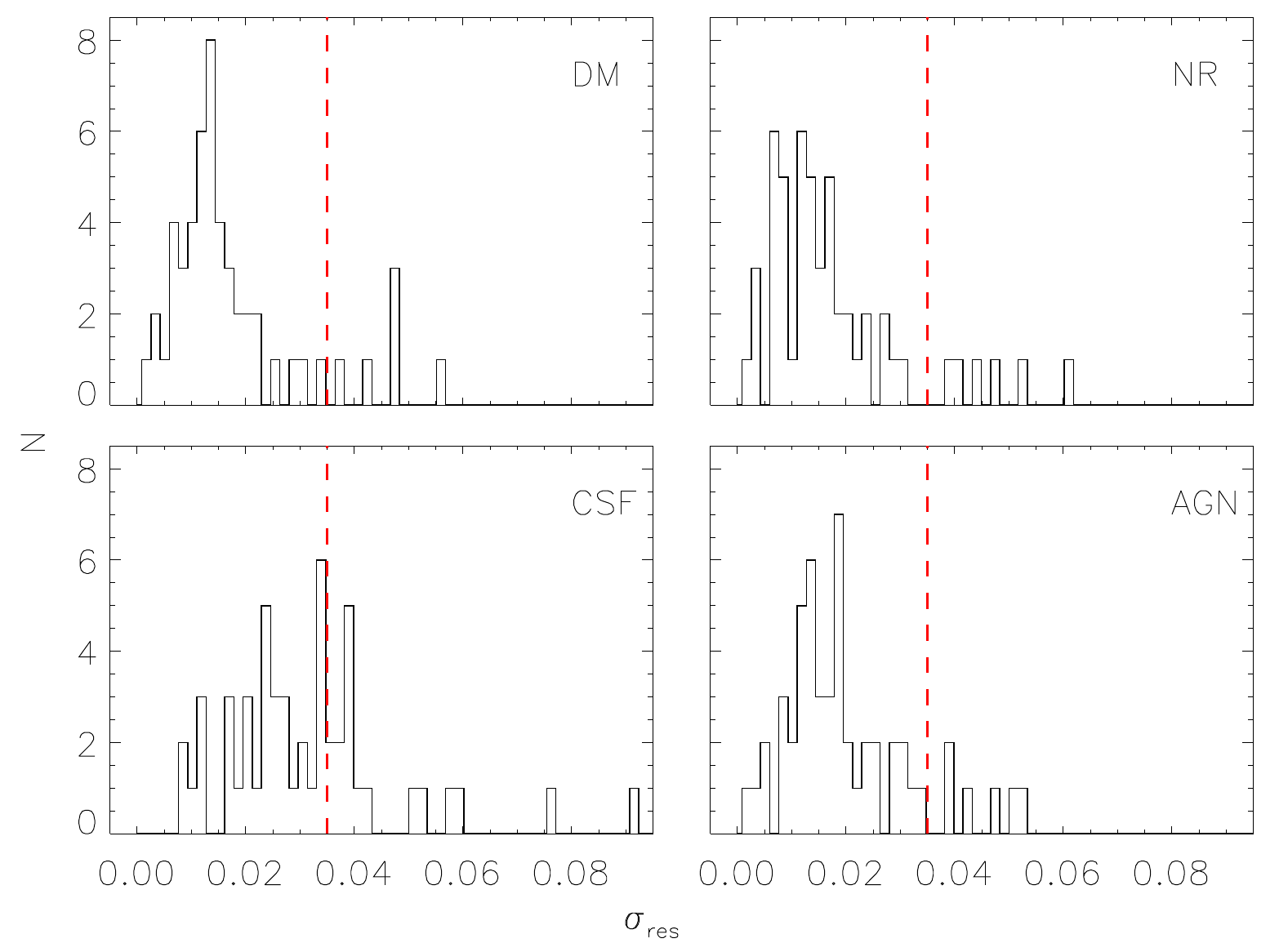}
\caption{Distribution of residuals, $\sigma_{\rm res}$, as defined in Eq.\ref{eq:residual} of the four physics and at redshift zero. The concentration is derived assuming the \simu\, radial range, $[0.06-1.27] \times R_{200}$, as defined in Section~\ref{sec:cmrr}. Acceptable clusters have $\sigma_{\rm res} < 0.035$ (values shown by a red dashed vertical line).}
\label{fig:cut}
\end{figure}
For all situations considered (varying ICM physics or radial ranges) and for both redshifts, we study the distributions of the residuals and notice that the majority of the objects have residuals below $ 0.035$ at $z=0$ (see Figure~\ref{fig:cut}) and $ 0.05$ at $z=0.25$. The residuals are significantly larger at $z=0.25$ especially for the NR physics. Furthermore, at fixed redshift, the outliers are often the same objects despite the change of the simulated set or the fitting radial range. Per each redshift, we identify these systems: nine clusters at $z=0$ and at $z=0.25$. The NR physics presents  larger deviations because accreted substructures requires more time to virialize and thermalize under the NR physical condition \citep{dolag.etal.09}.

In the following, we decide to always exclude  these identified systems and any other existing outliers (defined as having  $\sigma_{\rm res} > 0.035$ at $z=0$ and $\sigma_{\rm res}> 0.05$ at $z=0.25$).  The number of the remaining clusters is denoted by $N$ and is listed in Tables 1--3. For each situation investigated, before deriving the $c-M$ relation, we check whether the excluded halos are spread across the mass range. When this does not happen (for example, if all objects below a certain mass disappear) we {do not compute the $c-M$ relation because of the significant reduction in the sample mass range.

\subsection{The Effect of the Radial-range Choice}
\subsubsection{Typical Theoretical and Observational Radial Range}
\label{sec:cmrr}
Previous works demonstrated that  the fitting radial range  has some level of impact on the 
derived NFW concentration. For example, \cite{rasia.etal.06} showed that the concentration of a single cluster varies from $\sim$5.3 to 7 (23\% variation) when the outer radius moves from the virial radius to 20--30\% of that radius.  Using their X-ray observations on A2717, \cite{gastaldello.etal.07} warned of deriving  the 
concentration parameter if the data do not extend beyond the measured scale radius. \cite{fedeli12} stressed more the consequences of the choice of the inner radius: fitting from 1\% of $R_{200}$ induces a higher concentration, especially in small systems, than starting from 5\% of $R_{200}$.

To evaluate the influence of the radial range assumed to fit the mass profile with the NFW formula, we begin by studying the \dmonly sample.
We first consider radial ranges that are `typically' used in theory and X-ray studies, and then we move to a more general discussion.

Our references for theoretical  works are \cite{neto.etal.07}, \cite{duffy.etal.08, duffy.etal.10} and Meneghetti \& Rasia (2013). Their radial range used to fit the NFW formula is between 5 \% and 100\% of the virial radius, where the innermost limit was set to satisfy the requirement on the numerical convergence (Section~3.1 and \citealt{power.etal.03}).  This interval  is equal to $[0.06, 1.27] \times R_{200}$. We label this typical simulation radial range as SIM.

The X-ray radial range is, instead, linked to the one used in
\cite{ettori.etal.10}. The authors provided the radial boundaries used
to compute the NFW best fit. The inner radius was constantly fixed to
50 kpc  to exclude the impact of the central galaxy on the density
profile,  thus avoiding the stellar component influence. In our
whole sample, this value is larger than $0.03 \times R_{200}$
fulfilling the  numerical convergency requirement. E10 also furnished two outer
radii, one associated with the surface brightness profile and the other
with the spectroscopic temperature profile. Per each cluster, we select
the minimum of the two and compare its values to the derived
$R_{500}$. The resulting mean ratio was equal to 0.6.\footnote{In
  \cite{ettori.etal.10} two different $R_{500}$ radii were reported
  according to two different derivations of the mass profiles. For
  our purpose, the variation between these is minimal, as we
  tested.}  Concluding, we label the radial range from 50 kpc to $0.6
\times R_{500}$\footnote{This value corresponds roughly to $0.4 \times
  R_{200}$ for our objects.}  as X-ray.
 Furthermore, to account for recent X-ray observations reaching
 more external regions \citep[e.g.][]{humphrey.etal.12,walker.etal.13}, we enlarge the outer boundary of the range
 where the analysis is carried out to 0.8 and $1 \times R_{500}$. In all the cases analyzed in Section 3, the results, however,
 vary by only a few per-cent with respect to the results from the original X-ray
 radial range.

Considering both the SIM, and the X-ray, radial ranges, we fit the
same DM cluster mass profiles, derive values of both NFW
concentration and mass, and fit the $c-M$ relation
(Equation (4)). For the $z=0$ sample  we provide in the first two rows of
Table~\ref{tab:cmdm} the 
 median values of the {\it posterior} distributions of the
  normalization, $c_0$, and slope, $\alpha$, as long as the
  rms values of the intrinsic-scatter, $\sigma_{\epsilon_{\log}}$,
  and of the statistical scatter, $\sigma_{\rm stat}$.
The two slopes are consistent while  the normalizations differ by
  more than 1$\sigma$: when a typical X-ray, radial range is used,
the normalization increases by ~10\% 
 with respect to the value of the theoretical radial range. 
Limiting the fit to inner parts of the mass profile produces an
increase of the halo concentration at all mass scales. This is due to
the combined effect of lowering both the internal and external boundaries. In this case, the innermost part of the profile weights very strongly. 
Indeed, within 100 $h^{-1}$ kpc almost all the NFW X-ray\, best
fitting parameters are larger than those obtained for the SIM, case.
Comparing the concentrations, the scale radii, and the two $R_{200}$ derived, 
we found that with respect to the SIM, radial range: (1) the
X-ray, $R_{200} (=r_s \times c)$ is only 4\% lower; (2) the
X-ray, scale radii are significantly lower ($\sim$13\%) and,
consequently, (3) the X-ray, concentrations are higher ($\sim$16\%). 
 In both cases, the intrinsic scatter is around 10\%
  with only a 3\% statistical scatter. The intrinsic scatter is
  comparable to the values derived from observational data (B07, E10).
  The statistical scatter is instead slightly smaller than that
  reported by B07 ($\sim 0.06$) and significantly smaller than that
  found by E10 in the analysis of their entire cluster sample.
  This disagreement is mostly due to the different methods
  used to derive the concentrations. In fact, the concentrations in
  E10 are obtained as best-fitting parameters that minimize the
  difference between the reconstructed temperature profile and the
  profile obtained from the spatially resolved X-ray
  spectroscopy. \citet{meneghetti.etal.10} showed that the
  concentration errors associated with this procedure are on average
  three times larger than those obtained via the direct NFW fit of
  the mass profile obtained by the forward method,
  i.e., by applying the equation of hydrostatic equilibrium on the
  fitted emission measure and temperature profiles. However, we note
  that the statistical and intrinsic scatters of the best-determined
 relaxed systems analyzed by E10 are 3\% and 7\%, thus
  in agreement with our results.

\begin{table}[h]
\centering
\caption{Concentration-Mass Best-fit Parameters with 1$\sigma$ Error of the DM Sample at Redshift $z=0$. Apart the \xray\  radial range fixed between 50 kpc and $0.6 \times  R_{500}$, all remaining radial ranges are defined in units of $R_{200}$, with \simu=$[0.06-1.27] \times R_{200}$.  N is the number of objects satisfying the condition on the NFW residuals ($\sigma_{\rm res} < 0.035$). }
\label{tab:cmdm}
\vspace{10pt}
    \begin{tabular}{|c|cccc|}
\hline 
 & \multicolumn{4}{|c|}{\dmonly}\\
&$N$&$\alpha$& $c_0$ & $\sigma_{\epsilon_{\log}}$, $\sigma_{\rm stat}$\\ 
\hline
\hline
\simu                        &43 & {\bf --0.17  $\pm$   0.04}  &   {\bf 3.70  $\pm$   0.14}  &     0.10, 0.03 \\
\xray                         &42 & -0.15  $\pm$   0.05  &   4.06  $\pm$   0.14  &    0.09, 0.03  \\
\hline
$0.06-0.4$             &43  &-0.25  $\pm$   0.07  &   3.85  $\pm$   0.16  &     0.07, 0.05  \\
$0.06-2.0$           &43 & -0.15  $\pm$   0.04  &   3.73  $\pm$   0.13  &     0.10,  0.02 \\
$0.03-1.27$           &42 &-0.13  $\pm$   0.04  &   3.82  $\pm$   0.14  &     0.10, 0.02  \\
$0.2-1.27$            &43 &-0.31  $\pm$   0.10  &   3.57  $\pm$   0.19  &     0.07, 0.06  \\
\hline
\end{tabular}
\end{table}

\subsubsection{Generalization of the adopted radial range}

\begin{figure}[!h]
\centering
\includegraphics[width=0.48\textwidth]{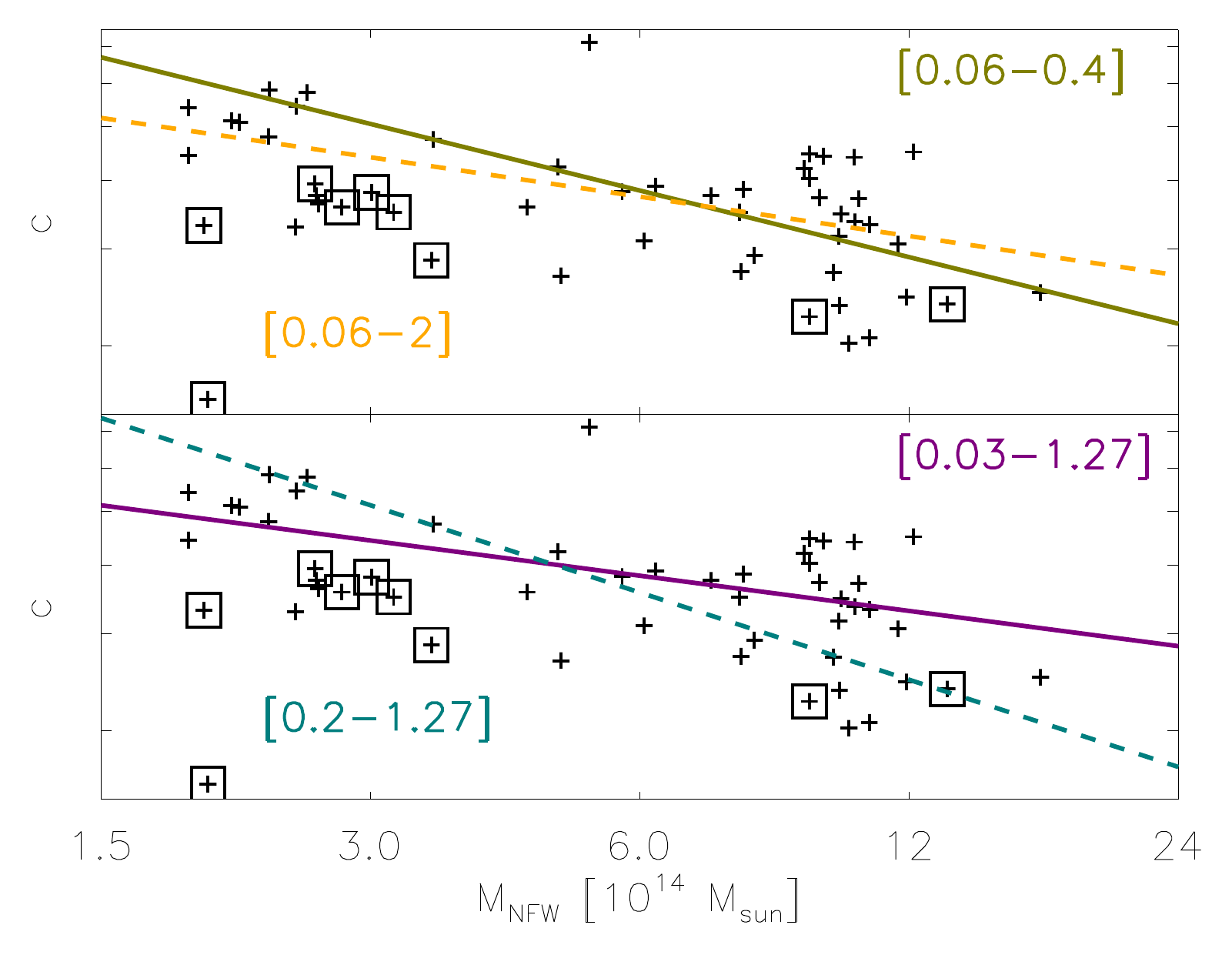}
\caption{Relation between NFW concentrations and masses computed for the {{\sc dm}} halos assuming the {{\sc sim}} radial range. Black squares indicate the 9 clusters excluded for their large residuals ($\sigma_{\rm res}>0.035$). The lines represent the fits to four different sets of concentration and mass values extracted by adopting four separate radial ranges (the 4 sets of points are omitted for clarity). With respect to the {{\sc sim}} radial range, we vary the  external radius from 0.4 (solid green line) to $2 \times R_{200}$ (dashed orange line) in the top panel, and the internal radius from 0.03 (solid purple line) to 0.2$\times R_{200}$ (dashed light blue line) in the bottom panel. }
\label{fig:dm_raggi2}
\end{figure}

To broaden the discussion about the radial ranges, we explore some
variation on the external and the internal radii with respect to the
SIM, radial range. The results are shown in
Figure~\ref{fig:dm_raggi2} and listed in  the lower part of
Table~\ref{tab:cmdm}. For clarity, we  show the cluster points only
for the SIM, radial range. In the top panel, the inner radius is
constantly equal to 0.06 $\times R_{200}$, while the external one
varies from 0.4 $\times R_{200}$ (solid 
 green 
line) to $2 \times R_{200}$ (dashed orange line).  The first
  number is close to the X-ray external radius ($0.6 \times R_{500}
  \sim 0.4 \times R_{200}$), while the second extends
  beyond the virial region. Carrying the analysis out to such larger
  radii (not reachable by optical and X-ray observations) aimed at
  testing the influence of the possible presence of filaments and
  accreting structures in the outskirts of clusters.  From this plot,
it appears that both slopes and normalization slightly increase when
the external radius becomes smaller and closer to the X-ray, limit .
 
 Even if the changes are very small, we tested the origin of this trend. We found that 
 halos with different masses react differently to changes of the radial fitting: smaller halos play the major role.
 For example, the parameters $\rho_s$ and $r_s$, of our most massive clusters ($M_{\rm true} > 10^{15} \hinv \msun$)\footnote{We indicate with $M_{\rm true}$ the true mass of the simulated objects within $R_{200}$. Notice that it might differ from the NFW masses reported in the various figures.} do not depend on the radial range chosen implying an averaged small variation on the NFW masses ($5$\%) and concentrations ($2$\%).  Our  smallest clusters ($M_{\rm true} < 1.5\times 10^{14} \hinv \msun$), instead, show a change in the scaling radius and in the normalization of $\sim 10$\% each. 
 The consequence  of using the narrowest radial range  is to
 produce a $5.5$\% larger concentration and $7.5$ \% smaller NFW mass
  with a shift not parallel to the SIM, $c_M$
   relation but almost orthogonal to it, thereby causing the overall change of the slope. We  infer that the radial range might play a more relevant role whenever the sample extends to small halos. 
 
The other scenario presented in the lower panel of the same Figure~\ref{fig:dm_raggi2} regards the change on the inner radius from 0.03$\times R_{200}$ (dashed  blue line) to $0.2 \times R_{200}$ (solid purple line), while the external radius is kept fixed at 1.27$\times R_{200}$.  With respect to the SIM radial range, excluding a significant portion of the inner profile (with a cut similar to that applied in weak-lensing studies) causes an  increase of the slope by almost a factor of 2.  
Extending the fit to inner regions  (i.e., to values comparable to
  the X-ray, inner radius), instead, slightly increases the
normalizations as previously noticed comparing the X-ray, and SIM
radial ranges. 
 The objects responsible for the difference between
  the two relations are the most massive clusters, thus at variance
  with the above considered case. Moving from the typical X-ray
  inner radius to the typical weak-lensing inner radius, concentrations
  and masses change by $8$\% and $5$\%, respectively,
  for the largest clusters, and only by $3$\%  and $1$\%
  for the smallest ones.

 In all the situations examined, however, the variation of the
  $c-M$ relation is not enough to explain the differences
  between the slopes of the simulated and observed relations of SA07
  and E10.


\subsection{The Effect of Baryons}
\label{sec:bar}

\begin{table*}
\centering
\caption{Concentration--Mass Best-fit Parameters with 1$\sigma$ Error of the Hydrodynamical Samples at Redshift $z=0$. Fitting radial
  ranges and symbols are as in Table~1. Left: results obtained from the total density profile; right: results related to the DM--only density profile.}
\label{tab:cmphy}
\vspace{10pt}
   \begin{tabular}{|c|cccc||cccc|}
\hline  
&\multicolumn{4}{c||}{\ovisc} &\multicolumn{4}{|c|}{DM in \ovisc} \\
&$N$&$\alpha$&$c_0$&$\sigma_{\epsilon_{\log}}$, $\sigma_{\rm stat}$&N&$\alpha$& $c_0$&$\sigma_{\epsilon_{\log}}$, $\sigma_{\rm stat}$\\
\hline   
 \simu                        & 43&   -0.17  $\pm$   0.05  &   3.98  $\pm$   0.17     &    0.12,    0.03&43& -0.18  $\pm$   0.05  &   3.66  $\pm$   0.16  &    0.11,   0.03\\
\xray                          & 42&   -0.19  $\pm$   0.05  &   4.06  $\pm$   0.19     &    0.12,   0.03&42& -0.19  $\pm$   0.06  &   3.86  $\pm$   0.17  &    0.11,   0.03\\
\hline
$[0.06-0.4]$               &43 &   -0.27  $\pm$   0.09  &   4.05  $\pm$   0.21   &    0.10,    0.05&41&-0.28  $\pm$   0.09  &   3.69  $\pm$   0.20  &    0.09,    0.05\\
$[0.06-2.0]  $            &41 &    -0.14  $\pm$   0.04  &   4.10  $\pm$   0.16   &    0.10,    0.02&42&-0.15  $\pm$   0.04  &   3.70  $\pm$   0.14  &    0.10,    0.02\\
$[0.03-1.27]$            &41  &   -0.14  $\pm$   0.04  &   4.02  $\pm$   0.17   &    0.12,    0.02 &41&-0.15 $\pm$   0.04  &   3.75  $\pm$   0.15  &   0.11,    0.02\\
$[0.21-1.27]$            &43  &   -0.32  $\pm$   0.12  &   3.92  $\pm$   0.22   &    0.08,    0.06 &43&-0.34  $\pm$   0.11  &   3.45  $\pm$   0.20  &    0.07,    0.06\\
\hline 
\hline
&\multicolumn{4}{c||}{\csf} &\multicolumn{4}{|c|}{DM in \csf} \\
&$N$&$\alpha$&$c_0$&$\sigma_{\epsilon_{\log}}$, $\sigma_{\rm stat}$&N&$\alpha$& $c_0$&$\sigma_{\epsilon_{\log}}$, $\sigma_{\rm stat}$\\
\hline   
\simu                           & 40& -0.21  $\pm$   0.05  &   4.76  $\pm$   0.21  &    0.11,   0.03& 43& -0.17  $\pm$   0.04  &   3.86  $\pm$   0.16 &    0.11,    0.03\\          
\xray                            &--   &  --           & --                                                     & --                        & 42 & -0.21  $\pm$   0.05  &   4.57  $\pm$   0.16  &    0.07,    0.03\\
\hline
$[0.06-0.4]$               &43& -0.30  $\pm$   0.05 &   5.53  $\pm$   0.26    &    0.06,    0.03 &43 & -0.30  $\pm$   0.08  &   4.04  $\pm$   0.21  &    0.07,   0.04 \\
$[0.06-2.0]  $             &36&  -0.20  $\pm$   0.04  &   4.86  $\pm$   0.18   &    0.10,    0.04&43 & -0.16  $\pm$   0.04  &   3.91  $\pm$   0.15  &    0.10,    0.02\\
$[0.03-1.27]$             &--& -- & --                                                                    & --                          &39 & -0.18  $\pm$   0.04  &   4.18  $\pm$   0.16  &    0.09,    0.02 \\
$[0.21-1.27]$             &43& -0.33  $\pm$   0.10  &   4.08  $\pm$   0.22   &    0.07,    0.06 &43 & -0.33  $\pm$   0.11  &   3.46  $\pm$   0.20  &    0.07,   0.06\\
\hline
\hline
&\multicolumn{4}{c||}{\agn} &\multicolumn{4}{|c|}{DM in \agn} \\
&N&$\alpha$&$c_0$&$\sigma_{\epsilon_{\log}}$, $\sigma_{\rm stat}$&N&$\alpha$& $c_0$&$\sigma_{\epsilon_{\log}}$, $\sigma_{\rm stat}$\\
\hline   
\simu                          &43&   -0.13  $\pm$   0.04  &   3.86  $\pm$   0.16      &    0.11,    0.03   &42&-0.14  $\pm$   0.04  &   3.57  $\pm$   0.15  &    0.11,    0.03 \\
\xray                           & 37&  -0.14  $\pm$   0.05  &   4.85  $\pm$   0.21      &    0.10,    0.03 &43&-0.16  $\pm$   0.05  &   3.45  $\pm$   0.15  &    0.11,     0.04\\
\hline 
$[0.06-0.4]$               &43&   -0.26  $\pm$   0.08  &   4.13  $\pm$   0.21       &    0.10,    0.05 &43&-0.25  $\pm$   0.09  &   3.66  $\pm$   0.20  &    0.09,    0.05\\
$[0.06-2.0]  $             &41&   -0.11  $\pm$   0.04  &   3.91  $\pm$   0.15       &    0.10,    0.02 &41&-0.12  $\pm$   0.04  &   3.62  $\pm$   0.14  &    0.10,    0.02\\
$[0.03-1.27]$             &24&   -0.14  $\pm$   0.04  &   4.21  $\pm$   0.18       &    0.11,    0.02 &40&-0.13  $\pm$   0.04  &   3.71  $\pm$   0.15  &    0.11,    0.02\\
$[0.21-1.27]$             &43&   -0.27  $\pm$   0.13  &   3.65  $\pm$   0.19        &    0.09,   0.06 &43&-0.32  $\pm$   0.12  &   3.31  $\pm$   0.20  &    0.08,    0.06 \\
\hline
\end{tabular}
\end{table*}

To evaluate the influence of baryons and of the baryonic physics on
the $c-M$ relation, we fit the NFW formula (Equation (2)) to the mass
profiles of the hydro simulations using all the radial ranges
discussed above (SIM, X-ray, and the four cases of
Figure~\ref{fig:dm_raggi2} and Table~1). Besides the total mass
profile, we consider in this section also the DM-only mass
profile,  that is, the mass profiles associated only with the dark-matter
  component within the hydrodynamical simulations.
Our results are summarized in
  Table~\ref{tab:cmphy} and described in the following.

\begin{figure}[!h]
\centering
\includegraphics[width=0.48\textwidth]{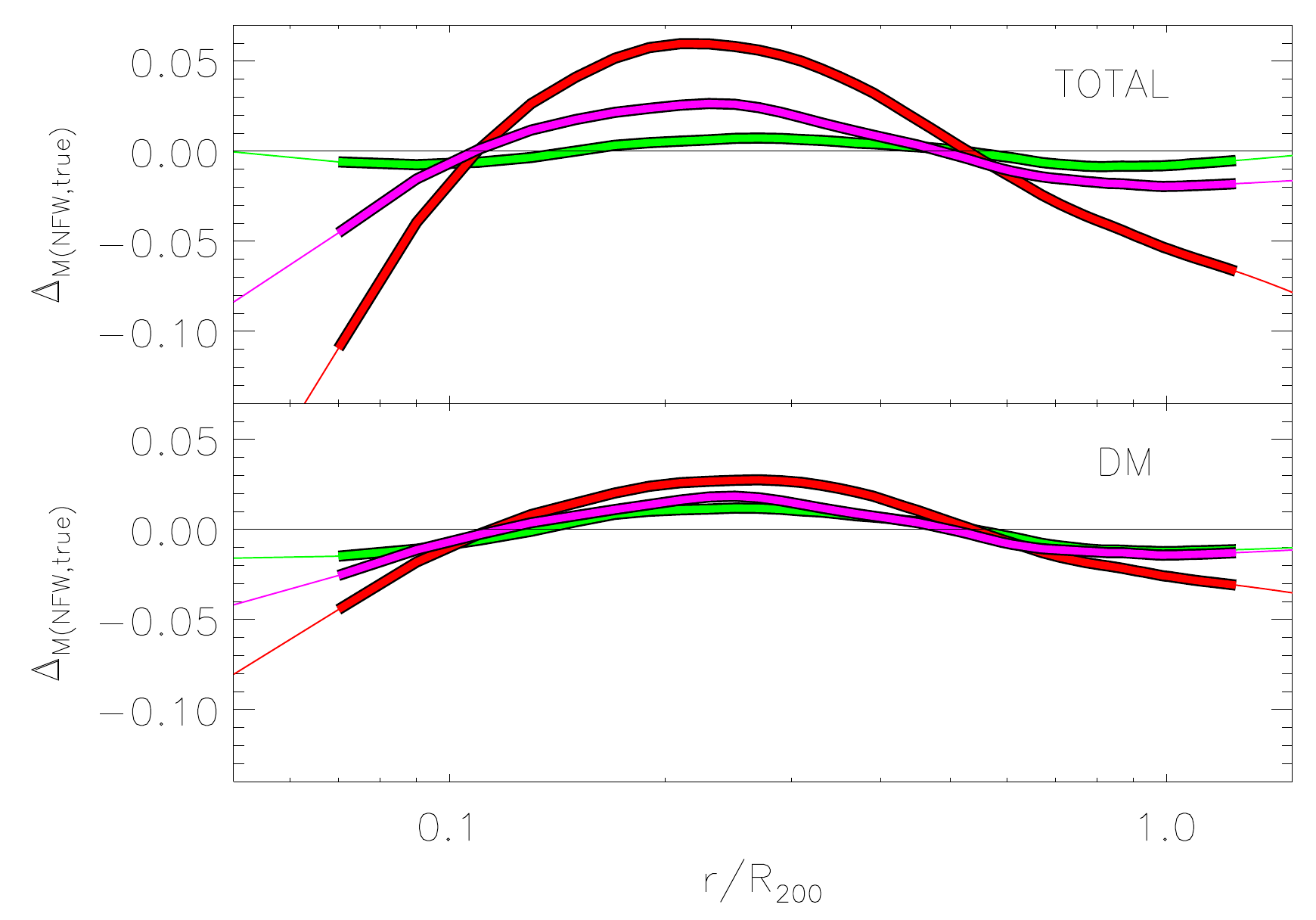}
\caption{Profiles of the residuals $\Delta_{M (\rm{NFW, true})}=(M_{\rm NFW}-M_{\rm true})/M_{\rm true}$ computed for the hydro simulations (NR in green, CSF in red, and AGN in magenta), assuming the {{\sc sim}} radial range, at redshift zero. In the top panel, the NFW fit is applied to {\it the total mass} profile, while in the bottom panel the mass profiles of the {\it only dark-matter particles} are considered.}
\label{fig:resi}
\end{figure}

\subsubsection{Deviation from the NFW formula}

 From Table~2, we note that the number of clusters, $N$, whose
  mass profiles are well fitted by the NFW expression (see
  Section~\ref{sec:profiles})  is  heavily reduced for the \csf
  sample. This is caused by the large deviation from the NFW profile
  especially in the innermost regions. This can be seen from Figure~\ref{fig:resi}, where we plot} the average profiles of the NFW-mass deviations from the true mass profile: $\Delta_{M (\rm{NFW,true})}=(M_{\rm NFW}-M_{\rm true})/M_{\rm true}$. The figure is built excluding the nine systems with large residuals. Even so, the \csf clusters (in red) largely depart from a pure NFW at the center.
The \ovisc clusters (in green) are very well represented by the NFW formula showing almost zero deviations at all radii. The \agn runs are halfway: including radiative cooling, as the \csf runs, they also have central profiles steeper than an NFW profile and than the \ovisc clusters. However, the presence of AGNs, which push a considerable amount of gas toward more external radii, reduces this effect with respect to the \csf simulations \citep{fabjan.etal.10}. 

The condensation of baryons in the center is very efficient in less
massive objects. Indeed, almost all our \csf groups with $M_{\rm
  true}<2 \times 10^{14} \hinv M_{\odot}$ have residuals
$\sigma>0.035$ when the fit is extended to small innermost limits
such as $0.03\times R_{200}$ and 50 kpc. The exclusion of all
these systems implies a drastic reduction of the mass range impeding
to robustly derive the $c-M$ relation. 
The statistics for these two cases being quite different, we do
  not report the corresponding values of the fitting parameters in Table~2.

 From the bottom panel of Figure~\ref{fig:resi}, we confirm that
  the NFW is an excellent description of the mass profile of the DM
  component within hydrodynamical simulations. Similar results for
  lower mass systems were obtained by \cite{duffy.etal.10}.  This
  feature persists in all studies obtained by varying the  radial range and the physics
  adopted, as witnessed by the value of $N$ in the right part of Table~2.

\subsubsection{Comparison with the DM Analysis: Slope}

The results of the hydrodynamical simulations (Table~\ref{tab:cmphy})
indicate that the slopes of the $c-M$ relations are always consistent
within 1$\sigma$ error with the DM slopes (Table~1). 
 The only major change lies in a more pronounced increase of the
  slope when either the internal or the external limit of the radial
  range is modified.  For instance, the difference in the AGN case is
  twice as much as the variation in the DM case when the external
  limit changes from the limit assumed in the SIM case to
  $0.4\times R_{200}$.
  The reason for this behavior can be understood from
  Figure~\ref{fig:agn}, where we plot the averaged density profiles
of the least massive (black line) and most massive (red line) clusters
renormalized to coincide at $0.5 \times R_{200}$. The profiles are not
exactly self-similar: least massive groups are on average more
concentrated than largest clusters.  In the bottom panel, we show how
much concentrations and NFW masses vary in relation to the external
radius. Most points are lying in a precise location because the
concentrations and masses of the two cases refer to the same potential
well as we are modifying only the  fitting radial range.  A
certain degeneracy between the parameters is, therefore, expected.
We find that our most massive systems ($M_{\rm true} > 10^{15}
h^{-1}M_{\odot}$),  on average, maintain the same concentration and mass despite the change of the radial range.
  On the other hand, the smallest
clusters ($M_{\rm true}<1.5 \times 10^{14} h^{-1} M_{\odot}$) show a
drop in concentration and an increase in mass  when extending the fit beyond the virial radius.
 
\subsubsection{Comparison with the DM Analysis: Normalization}
  
\begin{figure}[!h]
\centering
\includegraphics[width=0.48\textwidth]{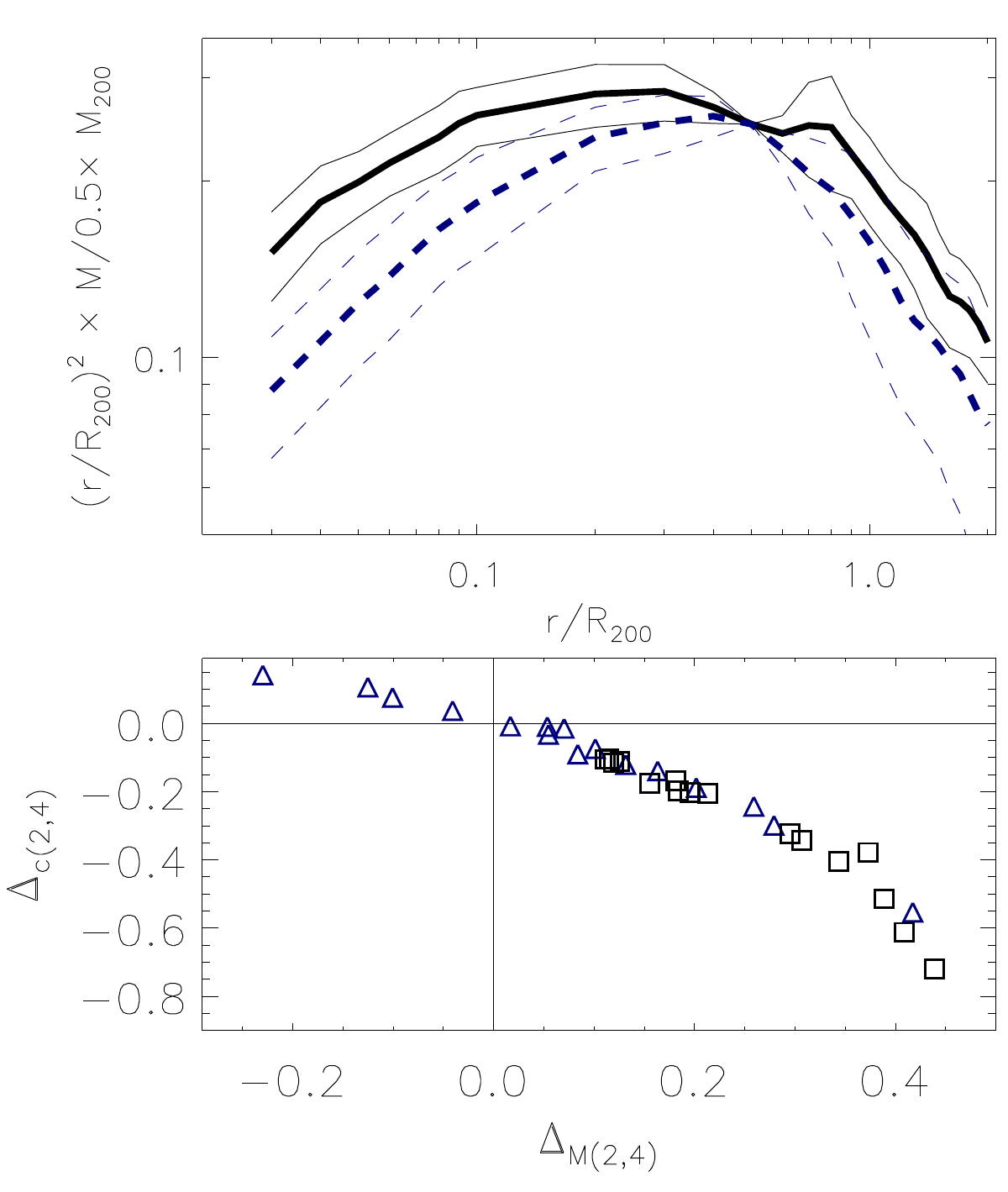}
\caption{Top panel: averaged mass profiles of the 17 least (black
  solid line) and the 17 most (light blue dashed line) massive clusters of
  the \agn sample, multiplied by the radius squared and normalized at
  $0.5\times R_{200}$. The thin lines represent the 1$\sigma$ range
  with respect to the mean value. {\it Quite clearly, the two profiles
    are not self-similar.} Bottom panel: difference between the NFW
  concentrations and masses: computed assuming two different radial
  ranges: [0.06-2]$\times R_{200}$ and $[0.06-0.4] \times
  R_{200}$. Naming the concentrations and masses as $c_2$ and $M_2$ in
  the first case and $c_4$ and $M_4$ in the second case, we plot the
  quantities $\Delta_{c_{2,4}}=(c_2-c_4)/c_2$ and
  $\Delta_{M_{2,4}}=(M_2-M_4)/M_2$. Black squares refer to the 17 least
  massive objects while light blue triangles refer to the 17 most
  massive ones. {\it The radial range chosen has a larger impact on the
    smallest clusters.} }
\label{fig:agn}
 \end{figure}

\begin{figure}[!h]
\centering
\includegraphics[width=0.48\textwidth]{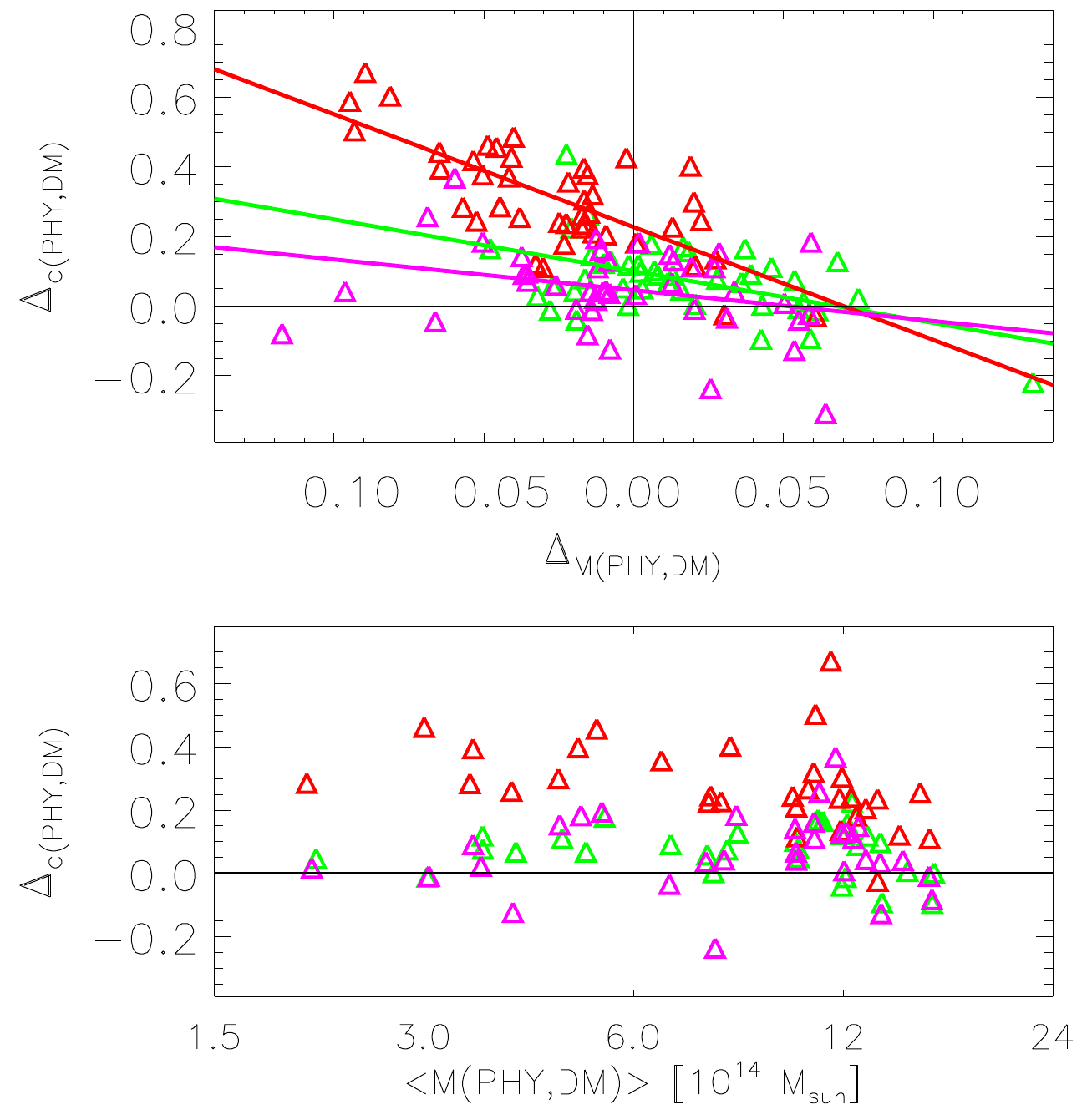}
\caption{Top panel: variation in the NFW concentrations and masses
  when the quantities are computed in the hydrodynamical simulations with respect to the \dmonly runs. Red, green, and magenta triangles represent {{\sc csf}}, {{\sc nr}}, and {{\sc agn}}\, results, respectively, assuming the {{\sc sim}}\, radial range. Bottom panel: the concentration shift is shown in function of the average NFW mass, calculated as the mean of the {{\sc dm}} and {{\sc phy}} masses.}
\label{fig:dc}
\end{figure}

  The largest difference between the $c$--$M$ relations predicted
   by hydrodynamical and by DM simulations lies in the normalization that shows 
 a systematic increase (Table~\ref{tab:cmphy})  of 5\% --
   10\% for the \ovisc runs, 15\% -- 50\% for the
   \csf runs, and of 25\% -- 40\% for the \agn runs.
 The radiative runs are those showing the largest deviations
 with respect to the DM case, especially when we reduce the outer
 limit of the radial range to 0.4 $\times R_{200}$. This result is
 shown in the top panel of Figure~\ref{fig:dc}, where, similarly to what
 is shown in the lower box of Figure~\ref{fig:agn}, we are reporting the
 variation in NFW concentrations and masses moving from the DM
 runs to the hydrodynamical simulations using the SIM radial range
 for $z=0$ systems. The \csf simulations produced a steeper relation
 and a higher normalization: the inclusion of baryons without a
 strong feedback induces an increase of the concentration at all mass
 scales (see lower panel in the figure). The effect is mitigated, if
 not completely canceled, when we include the strong feedback produced
 by the AGNs.  From the bottom panel, we can evince that the slope of
 \agn and \ovisc is very similar to the \dmonly one (both sets of
 green and magenta points are parallel to the \dmonly black horizontal
 line). Finally, we observe that the deviations in both concentration
 and mass are correlated (top panel), implying that even if the
 simulated underlying physics is changed, the  cluster mass profiles are not amply modified.

\subsubsection{Comparison with Previous Theoretical Works}

\cite{rudd.etal.08}, studying the influence of baryons on the matter
power spectrum, analyzed how baryons affect density profiles and
concentrations. Their results are very similar to ours: NR
simulations have a concentration boosted by 5\%-10\%  while
simulations with gas cooling, star formation, and SN feedback present
a larger difference (between 20\% and 50\%). Comparable values have
been found by \cite{deboni.etal.12}, 
 who also measured small variations
  associated 
  with the specific dark energy model considered.  
  
  Our results on the
  similarity between the concentration--mass relations for \agn and
  \dmonly simulations confirm the results found by
  \cite{duffy.etal.10} for the largest halos in their simulations ($M
  \sim 10^{14} \hinv \msun$) and extend to more massive clusters.

\subsubsection{Comparison with Previous Observational Works} 
 
 The comparison with the three reference X-ray analyses by B07,
  SA07, and E10, is carried out using the three scaling relations
  re-derived by \cite{fedeli12}.   
  The
  original papers adopted a $\Lambda$CDM cosmology with $h=0.7$ and
  $\Omega_M=0.3$. The first two works reported their measurements to
  the virial radius, the latter to $R_{200}$, and all assumed various
  modeling for the redshift evolution. In order to homogenize the
  results presented in the above observational analyses,
  \cite{fedeli12} adopted the following approach: (1) extrapolated
  all the results to the overdensity $\Delta_c=200$ (from
  \citealt{hu&kravtsov}); (2) assumed the best-fitting cosmological
  model from the {\it WMAP-7} analysis \citep{komatsu.etal.11}, similar to
  that assumed by our simulations; (3) removed any redshift
  dependence of the relation, and (4)
  chose a fixed pivot point at $5 \times 10^{14} \hinv \msun$, as
  we do. 

We refer to the X-ray papers for a complete description of the samples, here, we summarize the salient points: \\
\noindent  B07 studied 39  regular objects with masses from $6\times 10^{12} \msun$ to $2\times 10^{15} \msun$. The core of their sample resides in  early-type galaxies and groups with 16 systems of their sample with masses below $10^{14} \msun$. To these they added the massive clusters from the sample of \cite{pointecouteau.etal.05} and \cite{vikh.etal.06}. Restricting their analysis to only clusters with mass above $10^{14} \msun$, B07 confirmed what was previously found by the latter works. \\
\noindent  SA07 used 34 massive halos having mass-weighted temperatures within $R_{2500}$ above 5 keV and showing regular X-ray isoflux contours and minimal isophote centroid variations. Most of the X-ray information is contained in $R_{2500}$ ($\sim 0.3 \times R_{200}$).\\
\noindent Finally,  E10 considered 44 very luminous nearby
clusters with a minimum mass of $10^{14} \msun$  and defined a
  subset of 11 low-entropy core systems as representative of regular objects.

 S07 and E10 selected their clusters on the basis of the X-ray luminosity. The X-ray
  analysis differs among the three works as it does the fitting of the
  NFW functional form: B07 fitted the dark matter plus stellar mass
  profile, SA07 used the total mass profiles, while E10 considered
  only the DM profile (subtracting the gas and excluding the central
  regions where the stellar component dominates). These choices partially explain the differences among the
  observational results. Another important component is the sample selection. For example, selecting only low-entropy systems within their sample, E10 found a slope 60\% lower than considering the full ensemble of clusters.

\begin{figure}[h]
\centering
\includegraphics[width=0.48\textwidth]{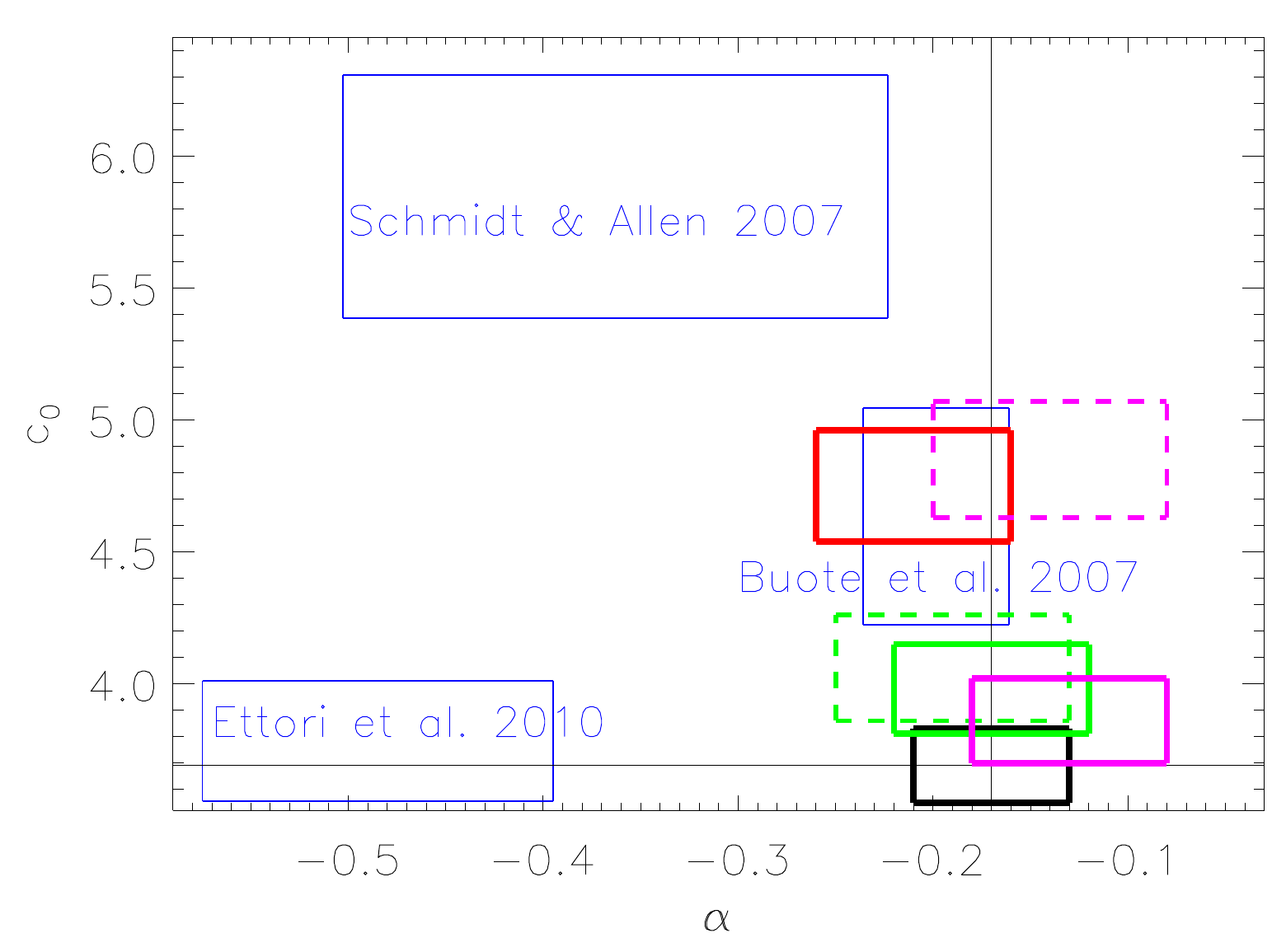}
\caption{Normalization, $c_0$, and slope, $\alpha$, of the $c-M$
  relations derived in the X-ray works present in literature (in blue)
  and in this paper (black, green, red, and magenta represent results
  derived from {{\sc dm}}, {{\sc nr}}, {{\sc csf}}, and {{\sc agn}}
  runs, respectively). Solid and dashed lines refer to results by fitting in the \simu~ and \xray~ radial ranges, respectively. The size of the boxes represents the $1\sigma$ variation.}
\label{fig:allplt}
\end{figure}

 Figure \ref{fig:allplt}
  confirms 
  that changing the gas physics does not lead to significant changes
  in the slope of the concentration--mass relation, using either the
  SIM or the X-ray radial range. The highest slope values,
  obtained for CSF clusters, are only 23\% higher than the
  DM slope and are significantly ($>1 \sigma$) smaller than the
  value found by E10 and only marginally consistent with SA07. On the other hand the slope derived by B07 is comparable to the one from our runs. As a note of caution, however,
  we remind that more than half of the observational sample analyzed has mass below our lower limit.

  The normalization is affected by a
larger degree by the change of the baryonic physics,  even if it
  never reaches the normalization of SA0, the only observational work
  based on fitting the NFW formula to the total mass profile.
The normalizations of the DM run are consistent with the value
derived by
\cite{ettori.etal.10}. 

 As a final remark, the intrinsic and the statistical scatters of
  the hydrodynamical simulations are almost identical to those of the
  DM runs (see also Section~3.3).

\vspace{0.3cm}
To conclude this section, we would like to point out that the $c-M$
 relation derived from the mass profiles of the DM
particles within the hydrodynamical simulations exhibits normalization and slope values very similar  among them and extremely close to those obtained from the pure $N$-body simulations. This result, joined with the fact that the DM profiles are always well represented by an NFW (lower panel of Figure \ref{fig:resi}), signifies that the best strategy to use in observations would be to compute the NFW concentration from the DM profile obtained by subtracting both the gas and stellar component from the total mass profiles \citep{newman.etal.13}.


\section{The observed c-M relation} \label{sec:cmo} 
\subsection{Synthetic X-Ray Analysis}

\begin{figure}[h]
\centering
\includegraphics[width=0.48\textwidth]{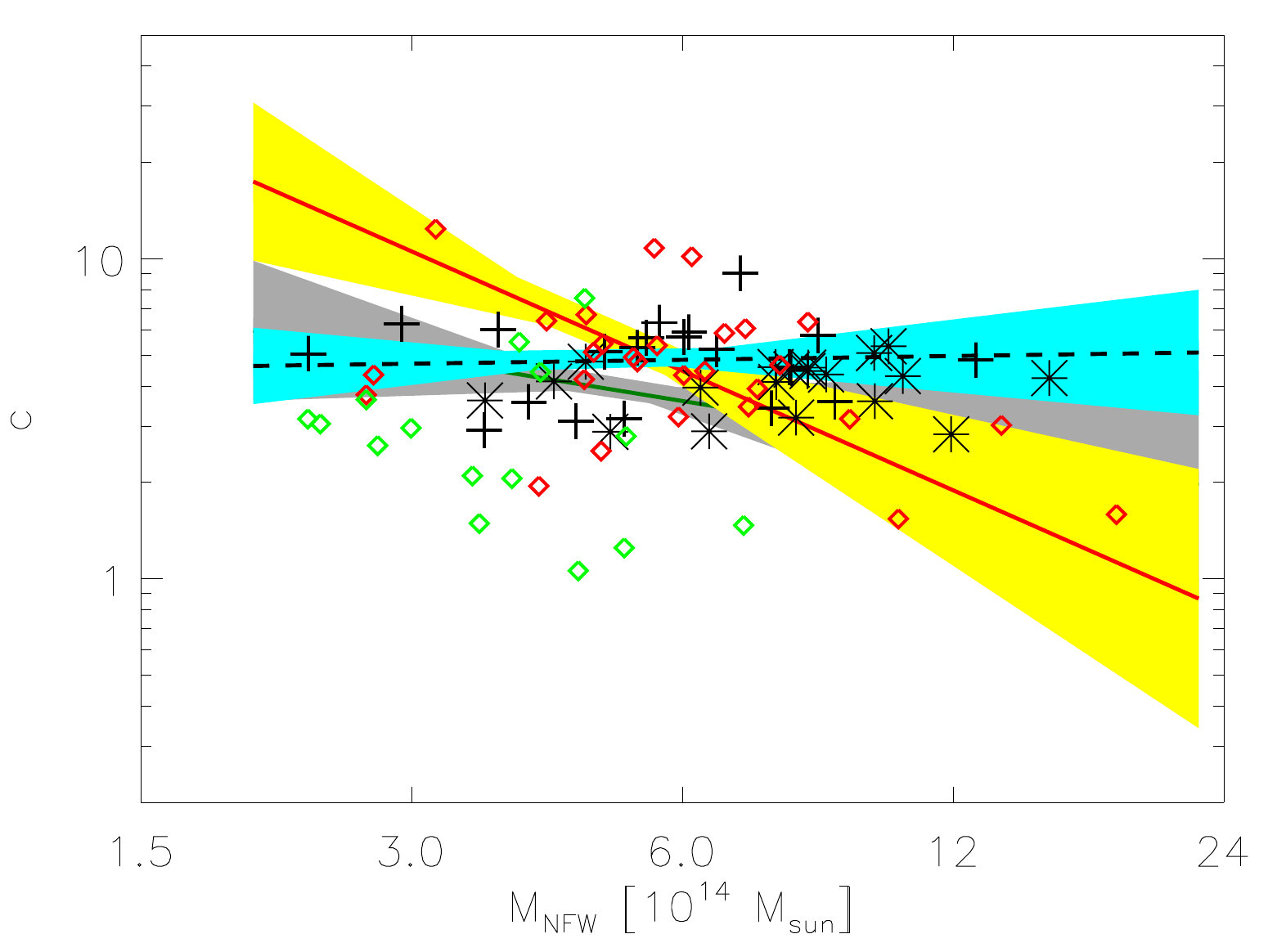}
\caption{Values of concentration and mass from the 3D intrinsic
  mass profiles ($c_i, M_i$; black asterisks), from the mass profiles
  derived by applying the hydrostatic equilibrium to the 3D gas
  density and 3D mass-weighted-temperature profile ($c_{he}, M_{he}$;
  black crosses), and from the X-ray mass profile of R12 ($c_x, M_x$;
  diamonds). In green we plot the values of {\it disturbed systems} as
  defined by \cite{rasia.etal.12b} while the red colors refer to the
  remaining objects. The $c_{he}-M_{he}$, $c_x,M_x$, and non-
    disturbed-cluster $c_x-M_x$ relations are shown with dashed black, solid
  green, and solid red line, respectively. The corresponding 1$\sigma$ ranges of
  variation are shown with the cyan, gray and yellow shaded regions.}
\label{fig:r12}
\end{figure}

 In this section we derive the $c-M$ relation  from the synthetic X-ray cluster catalog presented in R12 and compare it with the respective intrinsic relation. To this purpose,
we plot three sets of points in Figure~\ref{fig:r12}:
\begin{itemize}
\item[-] $c_i$, $M_i$ are the {\it intrinsic} parameters (black asterisks). The values are those of the 20 massive \csf clusters in common with the  synthetic X-ray catalogue and analyzed using the \simu\, radial range at $z=0.25$.
\item[-] $c_{he}$, $M_{he}$ are derived from the 3D mass profiles
  obtained by including in the hydrostatic equilibrium equation the
  true 3D gas density profile and the true 3D  mass-weighted
  temperature profile. These values are similarly obtained assuming
  the SIM, radial range (black crosses).  The
    $c-M$ relation and its 1$\sigma$ variation are shown
    by the black dashed line and cyan shaded region. 
\item[-] $ c_X$, $M_X$ are the values derived from the X-ray analysis. 
In R12, the X-ray mass profiles have been derived following the {\it forward} method \citep{vikh.etal.06,meneghetti.etal.10}: two fitting formulae have been adopted to fit the surface brightness and temperature profiles. The analytic best fits have been de-projected, and the mass calculated under the assumption of spherical symmetry and hydrostatic equilibrium. 
We adopted the mass profiles therein derived to fit the NFW formulae  using Equation (~\ref{eq:massnfw}) and Equation (\ref{eq:rhos})  in the radial range probed by the X-ray analysis performed in R12.
In this setting, the errors on the masses are taken directly from the
X-ray analysis output. They typically span from 10\%-12\% in
  the most central bins to $\sim 5$\% close to $R_{500}$, a value 
  similar to those in the literature \citep[E10]{vikh.etal.09}.  
The NFW concentrations and masses of the fitting profiles that agree within 1$\sigma$ with
the X-ray hydrostatic mass profiles  are shown in the figure with diamond symbols. The $c_X-M_X$ relation and its 1$\sigma$
  scatter are shown with dark green and gray shaded regions, respectively. In
  green, we plot the results for our {\it disturbed} systems as
  defined in \cite{rasia.etal.12b}.\footnote{See figures in that paper's
    Appendix for the cluster images.} Furthermore, we show in red
  results for the remaining systems,
  along with the corresponding 1$\sigma$ errors on concentration
  and NFW mass. The best fit to their $c-M$ relations is plotted
  with the solid red line. The shaded yellow region represents the
  1$\sigma$ range of variation.
  As found by SA07, the NFW formula is generally a good description of our non-disturbed X-ray data: only one system presents a residual above 0.035. 
  \end{itemize}

   The concentration--mass relations obtained by following
    different methods have {\it significantly} different slopes, ranging
    from $\alpha =0.04 \pm 0.31$ in the case of ($c_{\rm he}-M_{\rm he}$) to
    $-1.31 \pm 0.57$ in the case of the ($c_X-M_X$) for regular-only systems. In this comparison,
    the intrinsic scatters ($\sim 15$\%), the statistical
    scatters ($\sim 2$\%), and the normalizations ($c_0\sim 5$-5.5)
    are similar. Including the {\it disturbed} objects we still find
    a negative slope, $\alpha=-0.45\pm 0.49$, with a decrease of the 
    normalization, $c_0=3.91 \pm 0.39$, and a doubling of the intrinsic
    scatter, $25$\%,  but no significant variation of the statistical scatter.

  We proceed by analyzing separately two effects induced by  (1) the
    assumption of the hydrostatic equilibrium and (2) the X-ray
    oriented approach and analysis.
    
\begin{figure}[h]
\centering
\includegraphics[width=0.48\textwidth]{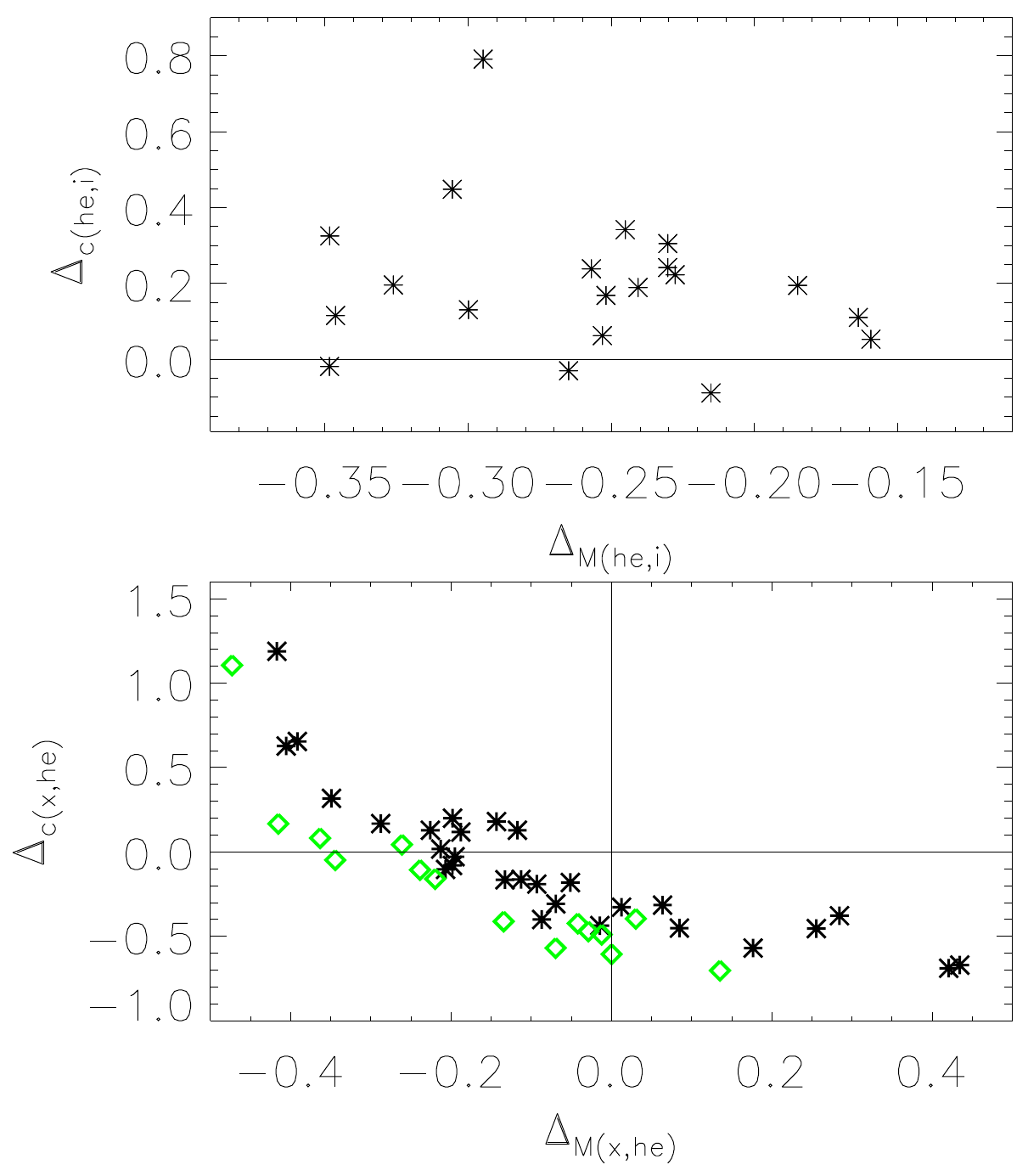}
\caption{Deviations in concentrations and in masses. Top panel:
  differences in the concentrations obtained from the intrinsic and from
  the hydrostatic approaches are plotted against the corresponding differences in mass.
  Bottom panel: the same as in the top panel, but when comparing
  results from the hydrostatic analysis and from the X-ray approach.
  Disturbed systems are shown with green diamonds. Horizontal and vertical lines show the case of no-variation {\it Irregular systems
    show a larger shift in both concentration and mass.}}
\label{fig:deltax}
\end{figure}

\subsection{Deviation Caused by the Hydrostatic Equilibrium Assumption}

 In the top panel of Figure~\ref{fig:deltax} we show how much concentrations and masses vary when we introduce the hydrostatic equilibrium assumption in our {\it intrinsic} analysis. In this case we define $\Delta_{c(he,i)}=(c_{he}-c_i)/c_i$ and $\Delta_{M(he,i)}=(M_{he}-M_i)/M_i$, where $i$ and $he$ indicate values shown in Figure~\ref{fig:r12} by asterisks and crosses, respectively. In average, the two masses show a difference of 25\%, with $M_{he}$ lower than $M_i$. This value is in agreement with the results on the masses presented in R12 (red line in their Figure~6), confirming that the NFW fitting procedure is not introducing any extra bias. The concentrations show a discrepancy of 20\% in the other direction without any specific correlation with either the mass of the systems, $M_i$, or the mass deviations, $\Delta_{M(he,i)}$.  
Consequentially,  we measure 30\%-40\% larger normalization for the $c-M$ relation $c_{he}-M_{he}$ when we assume the hydrostatic equilibrium, but we do not witness any change either in slope or in scatter  \cite[see also][]{lau.etal.09}. 

The phenomenon emphasizes that the lack of hydrostatic equilibrium in each system increases with the radius \citep{rasia.etal.04,rasia.etal.06,lau.etal.09,rasia.etal.12,khedekar.etal.12,battaglia.etal.12}. The profiles of $M_{he}$ are closer to the true mass profile at the center and diverge more in the external region,  automatically producing a more concentrated profile. The effect, however, is not sensible to the mass of the object, as we deduce from the fact that the slope of the $c-M$ relation does not vary. This consideration confirms the results by \cite{piffaretti&valdarnini}, who, using a more numerous sample (above 100 objects) and a larger mass range, found a negligible dependence on the mass of the hydrostatic-mass bias.

\subsection{Deviation Caused by the X-Ray Approach}
The outcomes differ when we move to compare the intrinsic hydrostatic values ($c_{he}$ and $M_{he}$) with the X-ray-derived one ($c_x$ and $M_x$). A significant increase of dispersion on both concentrations and masses  is present: most  clusters at high (low) mass have a lower (higher) concentration producing a clear steepening of the relation.
The deviations $\Delta_{c(x,he)}=(c_x-c_{he})/c_{he}$ and $\Delta_{M(x,he)}=(M_x=M_{he})/M_{he}$ are shown in the bottom panel of Figure~\ref{fig:deltax}.

The majority of the NFW masses from the X-ray data present a further reduction, with a median value of --12\%. The extra drop is caused by the temperature bias affecting this sample: the spectroscopic X-ray temperatures are lower than the mass-weighted ones because of dishomogeneity in the temperature distribution (R12).

The X-ray concentrations are significantly below the values derived from the 3D intrinsic profiles. The median deviation for the whole sample is --16\% and shows dependence on the X-ray morphology.
In Figure~\ref{fig:deltax}, we mark with green diamonds the systems belonging to the {\it disturbed} class.
The average concentration deviation  for the sub-sample of disturbed systems reaches --40\%.

\begin{figure}[h]
\includegraphics[width=0.48\textwidth]{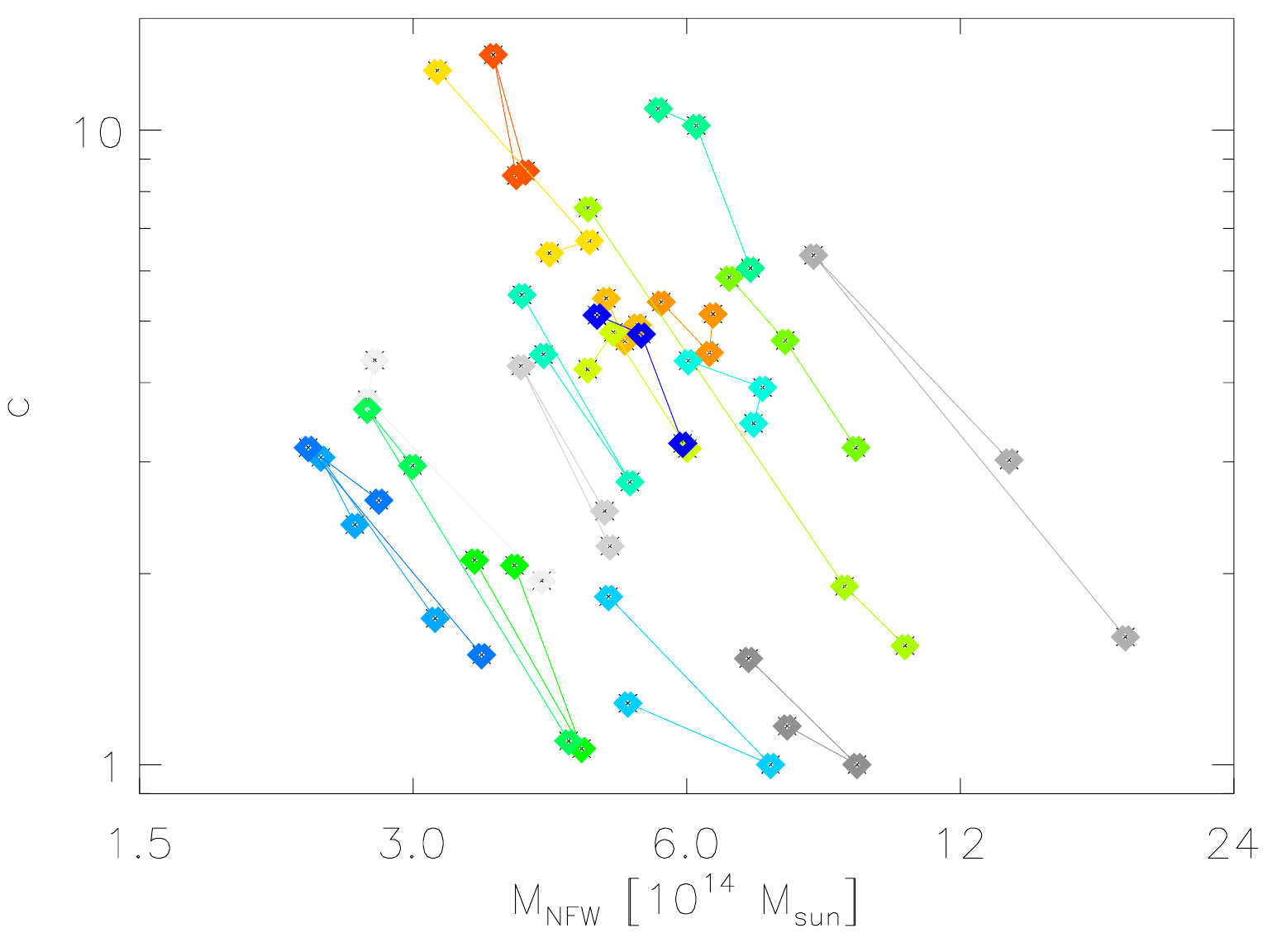} 
\includegraphics[width=0.48\textwidth]{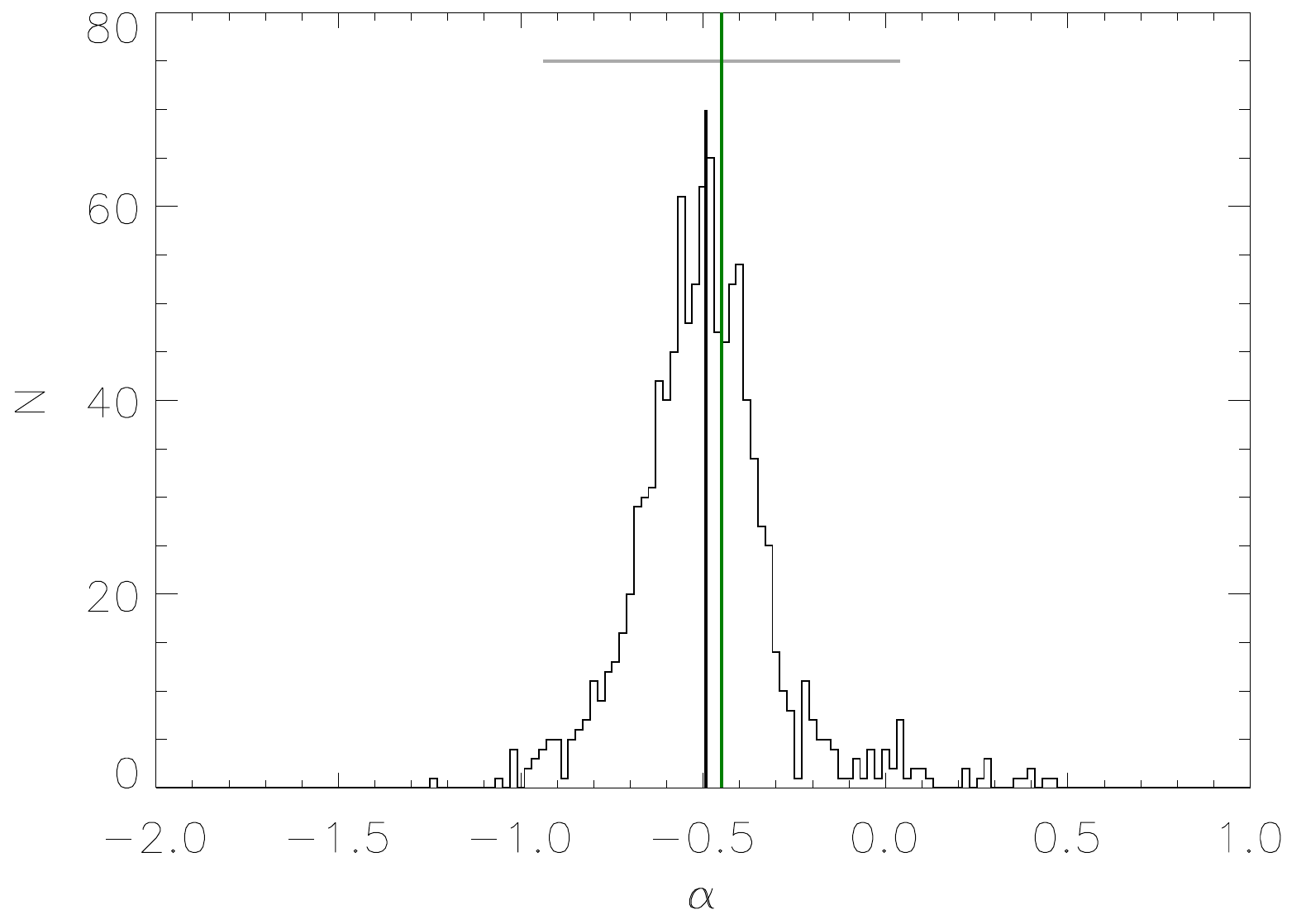} 
\caption{Top panel: the concentrations and the masses of the 
    synthetic X-ray catalogue. Each color refers to an individual
  cluster whose 3 projections are connected by a line.  Bottom panel:
  Distribution of the slope of the concentration-mass relation for
  1,000  re-samplings of the 20 clusters produced by randomly selecting one of their line-of-sight projection.
Black and green vertical lines
  correspond respectively to the median value of the distribution and the slope
  derived from the X-ray sample with 1$\sigma$ error shown with the
  gray horizontal line. {\it This plot demonstrates that there is less
    than 4\% of probability of obtaining a slope equal to or smaller
    than the intrinsic value (equal to $ 0.04$).
}}
\label{fig:cm_color}
\end{figure}

\subsection{Testing Degeneracy between Parameters}

The steepening of the relation might be a spurious effect originating from the fact that the 60 points are not completely independent; each single cluster is, instead, represented three times. Our concern is instigated by our previous results (Figures ~\ref{fig:agn} and \ref{fig:dc}), suggesting degeneracy between NFW masses and concentrations once they are derived from the similar potential wells using slightly different procedural settings (either radial range or simulation set or, in this context, separate projections).

In the top panel of Figure~\ref{fig:cm_color}, the set of points $c_x,
M_x$ are plotted with a choice of color that helps the identification
of the three projections of the same object.  Notice that almost
  always, with only two exceptions, clusters are classified as disturbed
  in all three projections. Clearly, this degeneracy is present.
What needs to be tested is whether the direction of the degeneracy is
parallel to the relation obtained by using only one projection or it
is orthogonal and, thus, steepens the relation.

To test this, 
we generate 2,000 possible combinations of our 20 objects. 
 Our resampling is built by  randomly selecting only one of the three available line-of-sight projections per cluster.
In this fashion, we are avoiding the degeneracy caused by having  `same-potential' systems in the sample. This approach is more indicated than some resampling methods such as {\it bootstrap} or {\it jackknife}  since their main assumption -- independency of the 60 measurements -- is definitely not satisfied in our case.  For each set, we derive the $c-M$ relation. The distribution of the $\sim 1000$ slopes with associated error less than 1 is shown in the bottom panel of Figure~\ref{fig:cm_color}. 

The X-ray slope of the $c_x-M_x$ relation (green vertical line)
is exactly in the middle of the distribution, meaning that the most
probable slope obtained by the 20 objects is almost the same as that
obtained from the enlarged sample of 60 images. 
From the histogram, we
find that there is less than 4\% probability to
find a slope as large as that obtained from the analysis based on
  hydrostatic equilibrium as a random realization from the
distribution obtained from the re-samplings of the 20 objects.
This test, therefore, reassures that we are not introducing an extra
bias from our sample.

\subsection{Understanding the Observational-like $c-M$ Relation}

Investigating, further, the reason for the steepening of the relation, we conclude that  many aspects, such as  projections, dynamical status, dynamical history, hydrostatic-equilibrium mass bias, temperature bias, environment, and the dependence of the biases on the mass of the systems, are contributing at the same time  without a clear driving factor.

The most extreme points are mainly responsible for this feature. They are identifiable in Figure~\ref{fig:r12} by having either concentration above 9 or mass above $10^{15} h^{-1} M_{\odot}$. The points with high concentration register a large underestimate of the hydrostatic-equilibrium masses and the NFW masses, justified by a considerable temperature bias. Since this bias increases with the radius, it leads to high concentration values. On the other hand, the points with low concentration and high NFW mass are disturbed objects located in rich environments. They do not show major substructures, but their X-ray emission remains high up to the outskirts, causing a flattening of the mass profile. Note that the X-ray enhancement is present only in one of the three projections; therefore, there is no flattening of the intrinsic mass profile.

Finally, even excluding these points, the slope of $c_x-M_x$ is
more than 1$\sigma$ larger than the intrinsic slope of $c_i-M_i$. The
explanation is mostly related to two factors noticed above:
restricting the external radial range produces an increase in the
slope, and most massive systems go in both directions of reducing the
concentration and increasing the mass and vice versa.

\section{Selection function}
Selecting the sample according to the  X-ray luminosity has been recognized as a source of bias in X-ray scaling relations \citep[e.g.,][]{nord.etal.08, pratt.etal.09,andreon.etal.11,andreon&hurn}. 
Indeed, clusters in a specific X-ray luminosity bin correspond to a broad range of masses due to the wide scatter of the luminosity--mass relation  
\citep{pratt.etal.09,stanek.etal.10,mantz.etal.10,allen.etal.12}. The correspondence between the theoretical  selection (based on mass) and the observational one (based on the X-ray emission) is, therefore, compromised.  At the same time, also the $c-M$ relation presents a significant dispersion:
galactic or small group masses can have a concentration that varies even by an order of magnitude \citep[e.g.,][]{rudd.etal.08, Ludlow.etal.12,bahe.etal.12,deboni.etal.12}. 

\begin{figure}[h]
\centering
\includegraphics[width=0.48\textwidth]{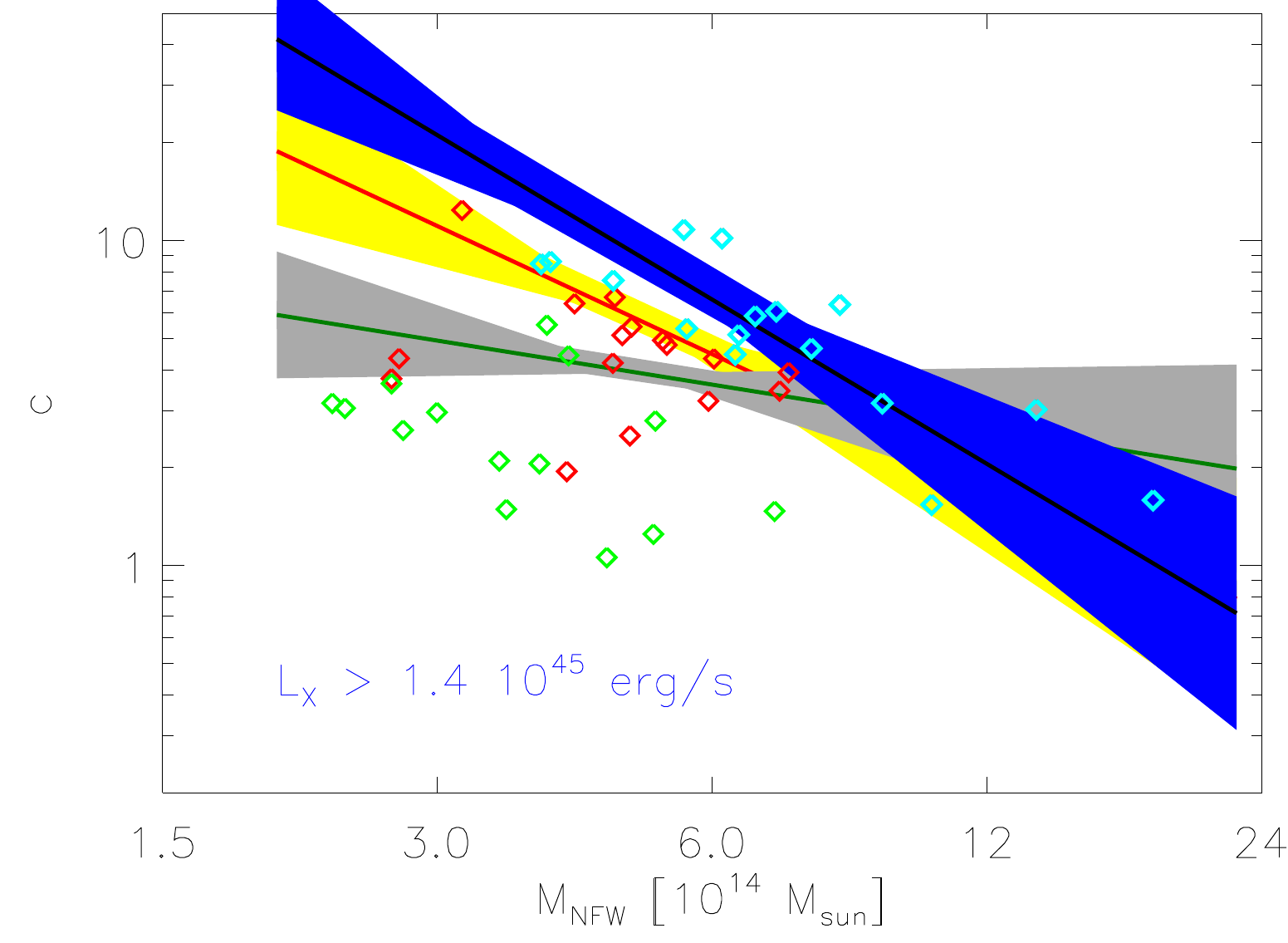}
\caption{The concentration-mass relation for the  synthetic X-ray
    catalogue. Clusters with [0.1-10] keV luminosity above 1.45
  $\times 10^{45}$ erg/s are shown in cyan. 
  The corresponding best-fitting relation is shown with a solid black
  line, while the blue region corresponds to the 1$\sigma$ variation. The
  colors of the other symbols, lines, and regions are the same as in
  Figure~7.}
\label{fig:flux}
\end{figure}

 Using our synthetic X-ray catalog, we investigate how the two
  scatters combine together when clusters are selected on the basis of
  their X-ray luminosity. In Figure~10, we plot our results where the
  blue shaded region is the $c-M$ relation derived from clusters more
  luminous than $1.45 \times 10^{45}$ erg s${^-1}$\footnote{The luminosity was computed in the [0.1-10] keV band.}. The increase in slope and
  normalization with respect  to the whole sample is quite clear:
  $\alpha=-1.68 \pm 0.55, c_0=8.93 \pm 2.07$, with an intrinsic scatter
  of about 12\%. The statistics on which this analysis is based
  is quite limited, given the relatively small number of simulated
  clusters. In order to overcome this limitation and strengthen our
  conclusions, we also consider the set of clusters identified at
  $z=0$ from a
  large cosmological box analyzed by \cite{deboni.etal.12}, obtained
  for a $WMAP-7$ cosmology. 
  The simulated volume of $(300 h^{-1} {\rm Mpc})^3$ was built with the Gadget-3 code
  evolving $(763)^3$ DM particles of mass $m \sim 3.7 \times 10^9 h^{-1} M_{\odot}$ and the same amount of gas particles.
  In Figure~11, we show the NFW concentrations and masses for $\sim$ -400 relaxed objects taken from that
  sample. We select three mass bins ($1.5<M_{\odot}/10^{14}<2$,
  $2<M_{\odot}/10^{14}<2/7$, $M_{\odot}/10^{14}>2.7$)
  to which we associate three soft-band X-ray luminosity bins
  obtained by adopting the luminosity--mass relation of the sample
  and using the X-ray [0.5--2] keV luminosities provided by \citet{deboni.etal.11}. The points
  associated with the three luminosity bins are shown with different
  colors. The horizontal solid red lines mark the median
  concentration of the luminosity bins while the median concentrations
  within the mass bins are shown with the black dashed lines. 
  Quite clearly, the luminosity selection biases the concentration
  toward higher values in each bin, the effect being larger in
  smaller halos, thus inducing a steepening of the slope of the
  $c-M$ relation. 

\begin{figure}[h]
\centering
\includegraphics[width=0.48\textwidth]{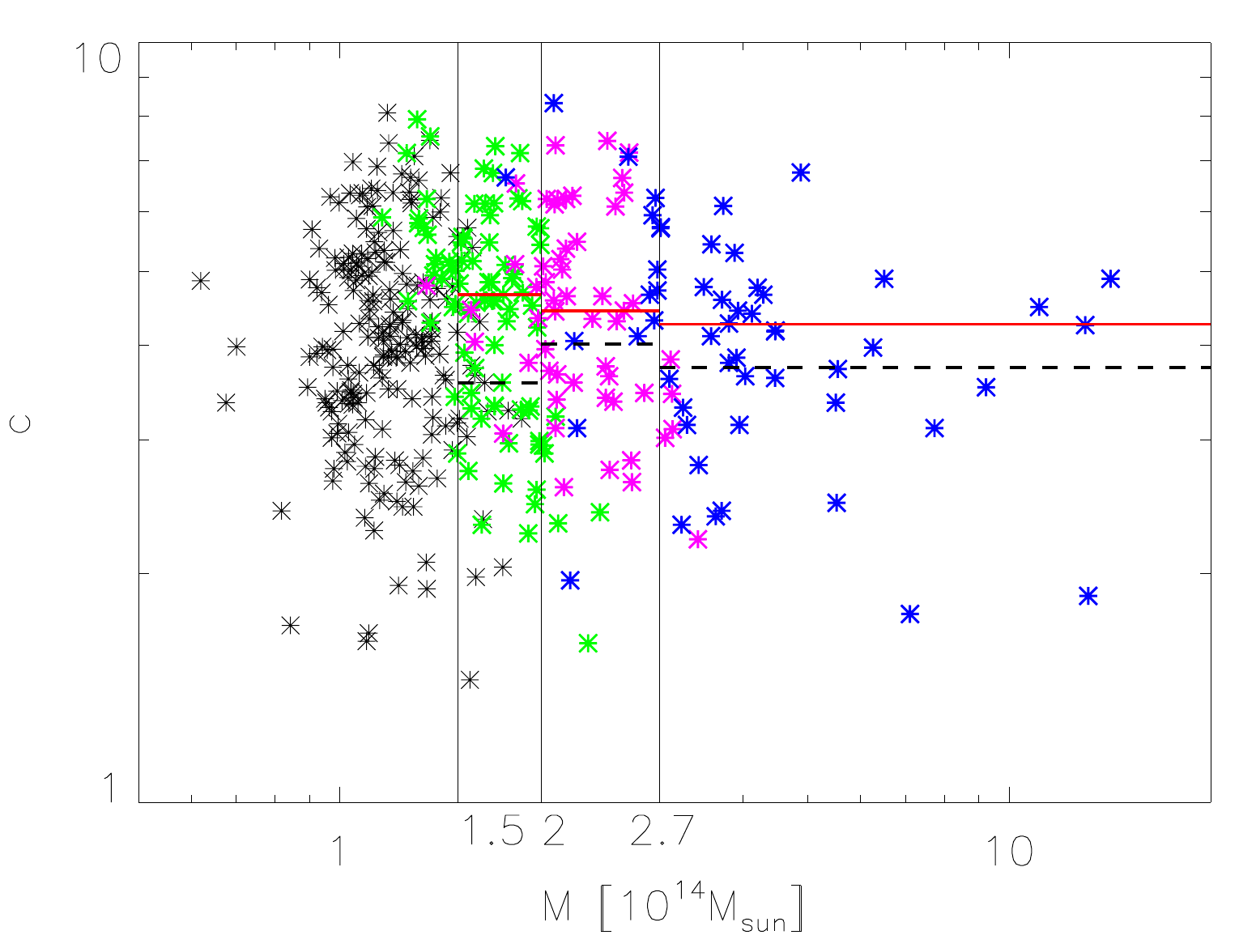}
\caption{The concentrations and masses from the {\sc wmap}
  cosmological box of \citet{deboni.etal.12} at redshift zero. Three
  vertical lines show the boundaries of three mass bins to which we
  associate three soft-band ([0.5-2] keV)  X-ray luminosity bins whose objects are colored in
  green, magenta, and blue from the least to the most luminous
  bin. Solid-red and dashed-black lines represent the median values of
  the concentrations obtained within the luminosity bins (points with the same color) and the mass
  bins (points within two vertical lines), respectively.}
\label{fig:flux}
\end{figure}

The findings derived from our synthetic catalog  might  explain the differences among the various observational $c-M$ relations. While E10 and SA07, with a $c-M$ slope around $-0.4$, are based on very luminous clusters, the samples of B07 and \cite{host&hansen}, with slopes closer to the theoretical one, involve also early-type galaxies and small groups that {\it were not selected} for their X-ray power. The normalization spread among these works is also in agreement with our findings about the differences caused by the change of the radial range: SA07 limited the information to very central regions while all the other works extended beyond $R_{2500}$.

\section{Conclusions} \label{sec:conclu} 

Recent claims about the discrepancy between the observed and simulated $c-M$ relation were declared real. Here, we tested if the difference
is instead induced by the two approaches followed to  derive the
values of NFW concentrations and masses. Specifically, we checked at
the influence of (1) the radial range adopted to perform the NFW fit
of the mass profile, (2) the baryonic-physics models included in
the simulation,  (3) the proper X-ray description, and (4) the
impact of the selection function. To accomplish the first two tasks, we
analyzed four sets of 52 clusters simulated four times, varying the
description of the ICM physics and the synthetic catalog of 60
$Chandra$-like X-ray images (R12). The four sets of physics
considered include dark matter (DM), no--radiative runs (NR), cooling, star formation and feedback by SN 
(CSF), and cooling, star formation, and feedback from AGNs (AGN).  The masses of the simulated clusters range from $\sim
  1.5 \times 10^{14} M_{\odot}$ to $\sim 2.5 \times 10^{15}
  M_{\odot}$. This allows us to carry out a direct comparison with
  observational works that analyze massive clusters, and to extend
  previous theoretical studies on the $c-M$ relation that
  typically focused on the mass scale of galaxies and galaxy groups.

\begin{itemize}
\item  {\it Radial range}. Using the DM set, we registered an
  increase in both slope and normalization of the $c-M$ relation when
  the most external radius considered during the NFW fit is
  reduced from outside the virial region ($ 2 \times R_{200}$) to the
  typical X-ray boundary ($0.6 \times R_{500} \sim 0.4 \times
  R_{200}$). Extending the fit to inner regions, instead, produces a
  higher normalization accompanied by a shallower slope.  The
    opposite trend is found when we exclude the innermost regions from the
    fit of the mass profile, as done in weak-lensing studies.
  All these differences, even if robust (intermediate choices of
  either internal or external radius reproduce the same trend of
  changes in slopes and normalizations), are mild, $\Delta \alpha < 50$\%
 and $\Delta c_0 < 10$\%.   They, however, become more pronounced once baryons are
  introduced ( $ \alpha$ and $c_0$ increases up to 100\% and 25\%, respectively).

\item {\it NFW fit and hydrodynamical simulations.}  The total mass profiles of the radiative simulations (\csf and {{\sc agn}}) show large residuals in the innermost region. The cause is the efficient condensation of baryons in the form of a stellar component that drastically steepens the total density profile. The \ovisc clusters at $z=0$ are, instead, well described by the NFW formula. 
For all physics and radial ranges, the results related to the DM component are always robustly similar.

\item {\it Baryon physics.} The relations derived from simulations
  with baryons have always a larger normalization than those from the
  \dmonly halos. 
We confirm that adiabatic contraction is quite
  important in radiative simulations,  a result that is in line
    with what was previously found in the analysis of less massive
    systems, although with a smaller amplitude
    \citep{rudd.etal.08,duffy.etal.10}. 
 Studying the influence of
  the radial range on hydrodynamical simulations, we noticed that
  smaller objects are more affected by the change of either internal
  or external radius.  The \csf set is characterized by a steeper and
  higher $c-M$ relation (increases around 30\% in both
  normalization and slope with respect to the \dmonly results in the
  \simu\, radial range). Both effects are strongly reduced when AGNs
  are included as a source of feedback. In this case, the slopes are
  much closer to the DM values, with variation of order of a few
  percent. The estimated differences for the  \ovisc data set are also less than 10\%.  Comparing the results of our intrinsic analysis
  with the observational ones, we found that the simulation slope
  overlaps with the values found only in \cite{buote.etal.07}, whose sample,
  however, extends to lower-mass objects. 
  
\item {\it X-ray $c-M$.} While the change of the radial range and the baryonic physics contribute only partially to explain the gap between the theoretical and observed $c-M$ slope, the analysis performed on our  synthetic X-ray catalog gave more insights. The hydrostatic-equilibrium assumption introduces a general reduction of the NFW masses (25\%) and an increase of the concentrations (20\%). The first result is explained by the HE mass bias. Its radial dependence justifies the second finding. The NFW masses and concentration derived from the  synthetic X-ray catalog have a huge dispersion (25\% as intrinsic scatter) with respect to the intrinsic results (15\% as intrinsic scatter). An increase in concentration corresponds to a decrease in mass and vice versa.
The slope derived from our X-ray sample is not influenced by the construction of our sample. The relation steepening is caused by the concomitant presence of multiple effects such as projections, environment, dynamical state, dynamical history, mass bias, temperature bias, and their dependences on the radius and on the mass of the system. Our morphologically regular objects show a more than twice steeper slope with respect to the entire sample. This might seem in contrast with the finding of E10, who reported a  shallower relation for their most relaxed systems. However, the selection of the two subsamples is based on different conditions: in our case regular systems have smaller third-order power ratio and centroid shift, while in E10 relaxed objects were defined as having lower central entropy level.

\item {\it Selection function} %
Utilizing the data from a full cosmological box \citep[the $(300 h^{-1} {\rm Mpc})^3$ volume simulated with a {\it wmap} cosmology at $z=0$ of][]{deboni.etal.12}, we verify that selecting clusters via their X-ray luminosity biases the $c-M$ relation toward both a higher normalization and slope with respect to the mass-selected sample. We obtain, indeed, that more X-ray  luminous clusters have, on average, higher concentration at fixed mass. %

\end{itemize}

Numerical works  present in the literature are based on many hundreds of halos. The effects of environment, dynamical status, and presence of substructures statistically cancel out when such alarge set of objects is considered. This is not the case when few tens of objects are selected as done in observational works. The comparison between the two samples, therefore, cannot be as straightforward.  From our analysis of the intrinsic profile, it emerged that smaller objects are more strongly affected by the choice of the radial range, and their mass profiles are more easily distorted by the mentioned effects. Furthermore, if groups or early-type ellipticals  are chosen for their X-ray power, luckily,  they will not represent the entire population of  similar-mass objects, but  will mostly be characterized by higher concentrations.


\appendix
In Table~\ref{tab:cm25} we report the normalizations and slopes of
the $c-M$ relations derived in the four sets of simulations at
$z=0.25$ adopting all the radial ranges presented above. The DM
profiles within the hydrodynamical simulation behave very similarly to the profiles from the DM set; therefore, we choose to not list them. The significant radii $R_{500}$ and $R_{200}$ are recomputed at $z=0.25$. Comparing the normalizations of the {{\sc sim}} radial range at $z=0.25$ with those at $z=0$, we notice that they are related to each other through: $c_0(z=0)=c_0(z=0.25)\times (1+z)^{0.4}$.

\begin{table*}
\centering
\caption{Concentration-mass best-fit parameters with 1 $\sigma$ error of the hydrodynamical samples at redshift $z=0.25$. Fitting radial ranges and symbols as in Table~1. Results obtained from the total density profile.}
\label{tab:cm25}
   \begin{tabular}{|c|c|cccc|}
\hline  
&&N&$\alpha$&$c_0$&$\sigma_{\epsilon_{\log}}$, $\sigma_{\rm stat}$ \\
\hline   
\hline
\dmonly &\simu          &43 & -0.10  $\pm$    0.04  &    3.42  $\pm$    0.15  &    0.11, 0.03       \\
&\xray                        &43 & -0.10  $\pm$    0.04  &    3.43  $\pm$    0.15  &    0.11, 0.04                      \\
\hline
&$[0.06-0.4]$             &43  &-0.17  $\pm$    0.07  &    3.51  $\pm$    0.20  &    0.10, 0.05                  \\
&$[0.06-2.0]  $           &43 & -0.08  $\pm$    0.04  &    3.47  $\pm$    0.14  &    0.10, 0.02                \\
&$[0.03-1.27]$           &43 &-0.08  $\pm$    0.04  &    3.56  $\pm$    0.15  &    0.11, 0.02                     \\
&$[0.21-1.27]$           &43 &-0.17  $\pm$    0.11  &    3.10  $\pm$    0.22  &    0.11, 0.06                 \\
\hline
\hline
\ \  \ovisc\  \   \  & \simu   & 34& -0.07  $\pm$    0.06  &    3.67  $\pm$    0.16  &    0.11, 0.03   \\
&\xray                             & 34&-0.07  $\pm$    0.06  &    3.66  $\pm$    0.16  &    0.10, 0.03  \\
\hline
&$[0.06-0.4]$               &34 & -0.23  $\pm$    0.16  &    3.71  $\pm$    0.21  &    0.10, 0.05    \\
&$[0.06-2.0]  $            & 34 & -0.05  $\pm$    0.05  &    3.68  $\pm$    0.15  &    0.10, 0.02    \\ 
&$[0.03-1.27]$            &32&-0.05  $\pm$    0.05  &    3.64  $\pm$    0.15  &    0.10, 0.02     \\
&$[0.21-1.27]$            &41 & -0.03  $\pm$    0.11  &    3.46  $\pm$    0.22  &    0.11, 0.06   \\
\hline 
\hline
\csf &\simu                    & 42&-0.18  $\pm$    0.04  &    4.26  $\pm$    0.19  &    0.11, 0.03\\
&\xray                           &42 &-0.18  $\pm$    0.04  &    4.27  $\pm$    0.20  &    0.11, 0.03\\
\hline
&$[0.06-0.4]$               &42& -0.27  $\pm$    0.04  &    4.85  $\pm$    0.27  &    0.06,  0.04\\
&$[0.06-2.0]  $             &36&  -0.14  $\pm$    0.05  &    4.38  $\pm$    0.20  &    0.11,  0.02\\
&$[0.03-1.27]$             &--& -- & -- &--                           \\
&$[0.21-1.27]$             &43& -0.21  $\pm$    0.07  &    3.42  $\pm$    0.21  &    0.11,  0.06\\
\hline
\hline
\agn&\simu                     & 43&-0.09  $\pm$    0.04  &    3.62  $\pm$    0.16  &    0.11, 0.03        \\
&\xray                           & 43&-0.10  $\pm$    0.04  &    3.62  $\pm$    0.16  &    0.11, 0.03        \\
\hline
&$[0.06-0.4]$               &43&-0.19  $\pm$    0.06  &    3.81  $\pm$    0.21  &    0.09, 0.04        \\
&$[0.06-2.0]  $             &43&-0.09  $\pm$    0.04  &    3.66  $\pm$    0.15  &    0.10, 0.02         \\
&$[0.03-1.27]$             &41&-0.09  $\pm$    0.04  &    3.94  $\pm$    0.16  &    0.10, 0.02        \\
&$[0.21-1.27]$             &43&-0.14  $\pm$    0.11  &    3.26  $\pm$    0.21  &    0.11, 0.06          \\
\hline
\end{tabular}
\end{table*}

\acknowledgments
\bigskip

{\bf Acknowledgment.}  We are grateful to the sharp and attentive
report by the anonymous referee.  E.R. thanks Cosimo Fedeli for sharing
his results on the concentration--mass relation and for discussions at
the early stage of this project  and Cristiano De Boni for
  making available the tables of the {\it wmap} cosmological box of
  \cite{deboni.etal.11} and \cite{deboni.etal.12}.  We thank Volker
Springel for providing us with the {\small GADGET--3} code.
Simulations have been carried out at the CINECA supercomputing centre
in Bologna, with CPU time assigned through ISCRA proposals and through
an agreement with University of Trieste.  We acknowledge financial
support by the following grants: National Science Foundation
AST-1210973, SAO TM3-14008X (issued under NASA Contract No.
NAS8-03060) , ASI-INAF I/023/05/0 and I/088/06/0, Marie Curie
Initial Training Network CosmoComp (PITN-GA-2009-238356) funded within
the European Commission's Framework Programme 7, PRIN-INAF09 project
``Towards an Italian Network for Computational Cosmology'',
PRIN-MIUR09 ``Tracing the growth of structures in the Universe'', and
INFN PD51.  E.R. and M.M. would like to thank the Michigan Center for
Theoretical Physics for supporting the collaboration.



\bibliographystyle{apj}

\begin{thebibliography}{97}
\expandafter\ifx\csname natexlab\endcsname\relax\def\natexlab#1{#1}\fi

\bibitem[{{Abadi} {et~al.}(2010){Abadi}, {Navarro}, {Fardal}, {Babul}, \&
  {Steinmetz}}]{abadi.etal.10}
{Abadi}, M.~G., {Navarro}, J.~F., {Fardal}, M., {Babul}, A., \& {Steinmetz}, M.
  2010, \mnras, 407, 435

\bibitem[{{Allen} {et~al.}(2011){Allen}, {Evrard}, \& {Mantz}}]{allen.etal.12}
{Allen}, S.~W., {Evrard}, A.~E., \& {Mantz}, A.~B. 2011, \araa, 49, 409

\bibitem[{{Anders} \& {Grevesse}(1989)}]{anders_grevesse}
{Anders}, E. \& {Grevesse}, N. 1989, \gca, 53, 197

\bibitem[{{Andreon} \& {Hurn}(2012)}]{andreon&hurn}
{Andreon}, S. \& {Hurn}, M.~A. 2012, ArXiv: 1210.6232

\bibitem[{{Andreon} {et~al.}(2011){Andreon}, {Trinchieri}, \&
  {Pizzolato}}]{andreon.etal.11}
{Andreon}, S., {Trinchieri}, G., \& {Pizzolato}, F. 2011, \mnras, 412, 2391

\bibitem[{{Bah{\'e}} {et~al.}(2012){Bah{\'e}}, {McCarthy}, \&
  {King}}]{bahe.etal.12}
{Bah{\'e}}, Y.~M., {McCarthy}, I.~G., \& {King}, L.~J. 2012, \mnras, 421, 1073

\bibitem[{{Balaguera-Antol{\'{\i}}nez} \& {Porciani}(2013)}]{balaguera.etal.12}
{Balaguera-Antol{\'{\i}}nez}, A. \& {Porciani}, C. 2013, \jcap, 4, 22

\bibitem[{{Baldi}(2012)}]{baldi12}
{Baldi}, M. 2012, Physics of the Dark Universe, 1, 162

\bibitem[{{Barnes} \& {White}(1984)}]{barnes.etal.84}
{Barnes}, J. \& {White}, S.~D.~M. 1984, \mnras, 211, 753

\bibitem[{{Battaglia} {et~al.}(2012){Battaglia}, {Bond}, {Pfrommer}, \&
  {Sievers}}]{battaglia.etal.12}
{Battaglia}, N., {Bond}, J.~R., {Pfrommer}, C., \& {Sievers}, J.~L. 2012,
  ArXiv: 1209.4082

\bibitem[{{Bhattacharya} {et~al.}(2013){Bhattacharya}, {Habib}, {Heitmann}, \&
  {Vikhlinin}}]{bhattacharya.etal.11}
{Bhattacharya}, S., {Habib}, S., {Heitmann}, K., \& {Vikhlinin}, A. 2013, \apj,
  766, 32

\bibitem[{{Blumenthal} {et~al.}(1986){Blumenthal}, {Faber}, {Flores}, \&
  {Primack}}]{blumenthal.etal.86}
{Blumenthal}, G.~R., {Faber}, S.~M., {Flores}, R., \& {Primack}, J.~R. 1986,
  \apj, 301, 27

\bibitem[{{Bonafede} {et~al.}(2011){Bonafede}, {Dolag}, {Stasyszyn}, {Murante},
  \& {Borgani}}]{bonafede.etal.11}
{Bonafede}, A., {Dolag}, K., {Stasyszyn}, F., {Murante}, G., \& {Borgani}, S.
  2011, \mnras, 418, 2234

\bibitem[{{Borgani} \& {Kravtsov}(2011)}]{borgani_kravtsov}
{Borgani}, S. \& {Kravtsov}, A. 2011, Advanced Science Letters, 4, 204

\bibitem[{{Bullock} {et~al.}(2001){Bullock}, {Kolatt}, {Sigad}, {Somerville},
  {Kravtsov}, {Klypin}, {Primack}, \& {Dekel}}]{bullock.etal.01}
{Bullock}, J.~S., {Kolatt}, T.~S., {Sigad}, Y., {Somerville}, R.~S.,
  {Kravtsov}, A.~V., {Klypin}, A.~A., {Primack}, J.~R., \& {Dekel}, A. 2001,
  \mnras, 321, 559

\bibitem[{{Buote} {et~al.}(2007){Buote}, {Gastaldello}, {Humphrey},
  {Zappacosta}, {Bullock}, {Brighenti}, \& {Mathews}}]{buote.etal.07}
{Buote}, D.~A., {Gastaldello}, F., {Humphrey}, P.~J., {Zappacosta}, L.,
  {Bullock}, J.~S., {Brighenti}, F., \& {Mathews}, W.~G. 2007, \apj, 664, 123

\bibitem[{{Chabrier}(2003)}]{chabrier03}
{Chabrier}, G. 2003, \pasp, 115, 763

\bibitem[{{Cui} {et~al.}(2012){Cui}, {Borgani}, {Dolag}, {Murante}, \&
  {Tornatore}}]{cui.etal.11}
{Cui}, W., {Borgani}, S., {Dolag}, K., {Murante}, G., \& {Tornatore}, L. 2012,
  \mnras, 423, 2279

\bibitem[{{Dalal} {et~al.}(2010){Dalal}, {Lithwick}, \&
  {Kuhlen}}]{dalal.etal.10}
{Dalal}, N., {Lithwick}, Y., \& {Kuhlen}, M. 2010, ArXiv: 1010.2539

\bibitem[{{De Boni} {et~al.}(2011){De Boni}, {Dolag}, {Ettori}, {Moscardini},
  {Pettorino}, \& {Baccigalupi}}]{deboni.etal.11}
{De Boni}, C., {Dolag}, K., {Ettori}, S., {Moscardini}, L., {Pettorino}, V., \&
  {Baccigalupi}, C. 2011, \mnras, 415, 2758

\bibitem[{{De Boni} {et~al.}(2013){De Boni}, {Ettori}, {Dolag}, \&
  {Moscardini}}]{deboni.etal.12}
{De Boni}, C., {Ettori}, S., {Dolag}, K., \& {Moscardini}, L. 2013, \mnras,
  428, 2921

\bibitem[{{Dolag} {et~al.}(2004){Dolag}, {Bartelmann}, {Perrotta},
  {Baccigalupi}, {Moscardini}, {Meneghetti}, \& {Tormen}}]{dolag.etal.04}
{Dolag}, K., {Bartelmann}, M., {Perrotta}, F., {Baccigalupi}, C., {Moscardini},
  L., {Meneghetti}, M., \& {Tormen}, G. 2004, \aap, 416, 853

\bibitem[{{Dolag} {et~al.}(2009){Dolag}, {Borgani}, {Murante}, \&
  {Springel}}]{dolag.etal.09}
{Dolag}, K., {Borgani}, S., {Murante}, G., \& {Springel}, V. 2009, \mnras, 399,
  497

\bibitem[{{Duffy} {et~al.}(2008){Duffy}, {Schaye}, {Kay}, \& {Dalla
  Vecchia}}]{duffy.etal.08}
{Duffy}, A.~R., {Schaye}, J., {Kay}, S.~T., \& {Dalla Vecchia}, C. 2008,
  \mnras, 390, L64

\bibitem[{{Duffy} {et~al.}(2010){Duffy}, {Schaye}, {Kay}, {Dalla Vecchia},
  {Battye}, \& {Booth}}]{duffy.etal.10}
{Duffy}, A.~R., {Schaye}, J., {Kay}, S.~T., {Dalla Vecchia}, C., {Battye},
  R.~A., \& {Booth}, C.~M. 2010, \mnras, 405, 2161

\bibitem[{{Eke} {et~al.}(2001){Eke}, {Navarro}, \& {Steinmetz}}]{eke.etal.01}
{Eke}, V.~R., {Navarro}, J.~F., \& {Steinmetz}, M. 2001, \apj, 554, 114

\bibitem[{{Ettori} {et~al.}(2010){Ettori}, {Gastaldello}, {Leccardi},
  {Molendi}, {Rossetti}, {Buote}, \& {Meneghetti}}]{ettori.etal.10}
{Ettori}, S., {Gastaldello}, F., {Leccardi}, A., {Molendi}, S., {Rossetti}, M.,
  {Buote}, D., \& {Meneghetti}, M. 2010, \aap, 524, 68

\bibitem[{{Fabjan} {et~al.}(2010){Fabjan}, {Borgani}, {Tornatore}, {Saro},
  {Murante}, \& {Dolag}}]{fabjan.etal.10}
{Fabjan}, D., {Borgani}, S., {Tornatore}, L., {Saro}, A., {Murante}, G., \&
  {Dolag}, K. 2010, \mnras, 401, 1670

\bibitem[{{Fedeli}(2012)}]{fedeli12}
{Fedeli}, C. 2012, \mnras, 424, 1244

\bibitem[{{Ferland} {et~al.}(1998){Ferland}, {Korista}, {Verner}, {Ferguson},
  {Kingdon}, \& {Verner}}]{ferland.etal.98}
{Ferland}, G.~J., {Korista}, K.~T., {Verner}, D.~A., {Ferguson}, J.~W.,
  {Kingdon}, J.~B., \& {Verner}, E.~M. 1998, \pasp, 110, 761

\bibitem[{{Gao} {et~al.}(2008){Gao}, {Navarro}, {Cole}, {Frenk}, {White},
  {Springel}, {Jenkins}, \& {Neto}}]{gao.etal.08}
{Gao}, L., {Navarro}, J.~F., {Cole}, S., {Frenk}, C.~S., {White}, S.~D.~M.,
  {Springel}, V., {Jenkins}, A., \& {Neto}, A.~F. 2008, \mnras, 387, 536

\bibitem[{{Gardini} {et~al.}(2004){Gardini}, {Rasia}, {Mazzotta}, {Tormen}, {De
  Grandi}, \& {Moscardini}}]{gardini.etal.04}
{Gardini}, A., {Rasia}, E., {Mazzotta}, P., {Tormen}, G., {De Grandi}, S., \&
  {Moscardini}, L. 2004, \mnras, 351, 505

\bibitem[{{Gastaldello} {et~al.}(2007){Gastaldello}, {Buote}, {Humphrey},
  {Zappacosta}, {Bullock}, {Brighenti}, \& {Mathews}}]{gastaldello.etal.07}
{Gastaldello}, F., {Buote}, D.~A., {Humphrey}, P.~J., {Zappacosta}, L.,
  {Bullock}, J.~S., {Brighenti}, F., \& {Mathews}, W.~G. 2007, \apj, 669, 158

\bibitem[{{Giocoli} {et~al.}(2012){Giocoli}, {Meneghetti}, {Ettori}, \&
  {Moscardini}}]{giocoli.etal.12}
{Giocoli}, C., {Meneghetti}, M., {Ettori}, S., \& {Moscardini}, L. 2012,
  \mnras, 426, 1558

\bibitem[{{Gnedin} {et~al.}(2011){Gnedin}, {Ceverino}, {Gnedin}, {Klypin},
  {Kravtsov}, {Levine}, {Nagai}, \& {Yepes}}]{gnedin.etal.11}
{Gnedin}, O.~Y., {Ceverino}, D., {Gnedin}, N.~Y., {Klypin}, A.~A., {Kravtsov},
  A.~V., {Levine}, R., {Nagai}, D., \& {Yepes}, G. 2011, ArXiv: 1108.5736

\bibitem[{{Gnedin} {et~al.}(2004){Gnedin}, {Kravtsov}, {Klypin}, \&
  {Nagai}}]{gnedin.etal.04}
{Gnedin}, O.~Y., {Kravtsov}, A.~V., {Klypin}, A.~A., \& {Nagai}, D. 2004, \apj,
  616, 16

\bibitem[{{Governato} {et~al.}(2012){Governato}, {Zolotov}, {Pontzen},
  {Christensen}, {Oh}, {Brooks}, {Quinn}, {Shen}, \&
  {Wadsley}}]{governato.etal.12}
{Governato}, F., {Zolotov}, A., {Pontzen}, A., {Christensen}, C., {Oh}, S.~H.,
  {Brooks}, A.~M., {Quinn}, T., {Shen}, S., \& {Wadsley}, J. 2012, \mnras, 422,
  1231

\bibitem[{{Grossi} \& {Springel}(2009)}]{grossi&springel}
{Grossi}, M. \& {Springel}, V. 2009, \mnras, 394, 1559

\bibitem[{{Gustafsson} {et~al.}(2006){Gustafsson}, {Fairbairn}, \&
  {Sommer-Larsen}}]{gustafsson.etal.06}
{Gustafsson}, M., {Fairbairn}, M., \& {Sommer-Larsen}, J. 2006, \prd, 74,
  123522

\bibitem[{{Haardt} \& {Madau}(2001)}]{haardt&madau01}
{Haardt}, F. \& {Madau}, P. 2001, in Clusters of Galaxies and the High Redshift
  Universe Observed in X-rays, ed. D.~M. {Neumann} \& J.~T.~V. {Tran}

\bibitem[{{Host} \& {Hansen}(2011)}]{host&hansen}
{Host}, O. \& {Hansen}, S.~H. 2011, \apj, 736, 52

\bibitem[{{Hu} \& {Kravtsov}(2003)}]{hu&kravtsov}
{Hu}, W. \& {Kravtsov}, A.~V. 2003, \apj, 584, 702

\bibitem[{{Humphrey} {et~al.}(2012){Humphrey}, {Buote}, {Brighenti}, {Flohic},
  {Gastaldello}, \& {Mathews}}]{humphrey.etal.12}
{Humphrey}, P.~J., {Buote}, D.~A., {Brighenti}, F., {Flohic}, H.~M.~L.~G.,
  {Gastaldello}, F., \& {Mathews}, W.~G. 2012, \apj, 748, 11

\bibitem[{{Kelly}(2007)}]{kelly.etal.07}
{Kelly}, B.~C. 2007, \apj, 665, 1489

\bibitem[{{Khedekar} {et~al.}(2013){Khedekar}, {Churazov}, {Kravtsov},
  {Zhuravleva}, {Lau}, {Nagai}, \& {Sunyaev}}]{khedekar.etal.12}
{Khedekar}, S., {Churazov}, E., {Kravtsov}, A., {Zhuravleva}, I., {Lau}, E.~T.,
  {Nagai}, D., \& {Sunyaev}, R. 2013, \mnras, 431, 954

\bibitem[{{Killedar} {et~al.}(2012){Killedar}, {Borgani}, {Meneghetti},
  {Dolag}, {Fabjan}, \& {Tornatore}}]{killedar.etal.12}
{Killedar}, M., {Borgani}, S., {Meneghetti}, M., {Dolag}, K., {Fabjan}, D., \&
  {Tornatore}, L. 2012, \mnras, 427, 533

\bibitem[{{Komatsu} {et~al.}(2011){Komatsu}, {Smith}, {Dunkley}, {Bennett},
  {Gold}, {Hinshaw}, {Jarosik}, {Larson}, {Nolta}, {Page}, {Spergel},
  {Halpern}, {Hill}, {Kogut}, {Limon}, {Meyer}, {Odegard}, {Tucker}, {Weiland},
  {Wollack}, \& {Wright}}]{komatsu.etal.11}
{Komatsu}, E., {Smith}, K.~M., {Dunkley}, J., {Bennett}, C.~L., {Gold}, B.,
  {Hinshaw}, G., {Jarosik}, N., {Larson}, D., {Nolta}, M.~R., {Page}, L.,
  {Spergel}, D.~N., {Halpern}, M., {Hill}, R.~S., {Kogut}, A., {Limon}, M.,
  {Meyer}, S.~S., {Odegard}, N., {Tucker}, G.~S., {Weiland}, J.~L., {Wollack},
  E., \& {Wright}, E.~L. 2011, \apjs, 192, 18

\bibitem[{{Kravtsov} \& {Borgani}(2012)}]{kravtsov&borgani12}
{Kravtsov}, A.~V. \& {Borgani}, S. 2012, \araa, 50, 353

\bibitem[{{Kwan} {et~al.}(2012){Kwan}, {Bhattacharya}, {Heitmann}, \&
  {Habib}}]{kwan.etal.12}
{Kwan}, J., {Bhattacharya}, S., {Heitmann}, K., \& {Habib}, S. 2012, ArXiv:
  1210.1576

\bibitem[{{Lau} {et~al.}(2009){Lau}, {Kravtsov}, \& {Nagai}}]{lau.etal.09}
{Lau}, E.~T., {Kravtsov}, A.~V., \& {Nagai}, D. 2009, \apj, 705, 1129

\bibitem[{{Lau} {et~al.}(2011){Lau}, {Nagai}, {Kravtsov}, \&
  {Zentner}}]{lau.etal.11}
{Lau}, E.~T., {Nagai}, D., {Kravtsov}, A.~V., \& {Zentner}, A.~R. 2011, \apj,
  734, 93

\bibitem[{{Ludlow} {et~al.}(2012){Ludlow}, {Navarro}, {Li}, {Angulo},
  {Boylan-Kolchin}, \& {Bett}}]{Ludlow.etal.12}
{Ludlow}, A.~D., {Navarro}, J.~F., {Li}, M., {Angulo}, R.~E., {Boylan-Kolchin},
  M., \& {Bett}, P.~E. 2012, \mnras, 427, 1322

\bibitem[{{Macci{\`o}} {et~al.}(2012){Macci{\`o}}, {Stinson}, {Brook},
  {Wadsley}, {Couchman}, {Shen}, {Gibson}, \& {Quinn}}]{maccio.etal.12}
{Macci{\`o}}, A.~V., {Stinson}, G., {Brook}, C.~B., {Wadsley}, J., {Couchman},
  H.~M.~P., {Shen}, S., {Gibson}, B.~K., \& {Quinn}, T. 2012, \apjl, 744, L9

\bibitem[{{Mantz} {et~al.}(2010){Mantz}, {Allen}, {Ebeling}, {Rapetti}, \&
  {Drlica-Wagner}}]{mantz.etal.10}
{Mantz}, A., {Allen}, S.~W., {Ebeling}, H., {Rapetti}, D., \& {Drlica-Wagner},
  A. 2010, \mnras, 406, 1773

\bibitem[{{Martizzi} {et~al.}(2012){Martizzi}, {Teyssier}, \&
  {Moore}}]{martinizzi.etal.12}
{Martizzi}, D., {Teyssier}, R., \& {Moore}, B. 2012, ArXiv e-prints

\bibitem[{{Meneghetti} {et~al.}(2010){Meneghetti}, {Rasia}, {Merten},
  {Bellagamba}, {Ettori}, {Mazzotta}, {Dolag}, \& {Marri}}]{meneghetti.etal.10}
{Meneghetti}, M., {Rasia}, E., {Merten}, J., {Bellagamba}, F., {Ettori}, S.,
  {Mazzotta}, P., {Dolag}, K., \& {Marri}, S. 2010, \aap, 514, A93

\bibitem[{{Navarro} {et~al.}(1996){Navarro}, {Frenk}, \& {White}}]{nfw96}
{Navarro}, J.~F., {Frenk}, C.~S., \& {White}, S.~D.~M. 1996, \apj, 462, 563

\bibitem[{Navarro {et~al.}(1997)Navarro, Frenk, \& White}]{nfw}
Navarro, J.~F., Frenk, C.~S., \& White, S. D.~M. 1997, \apj, 490, 493

\bibitem[{{Neto} {et~al.}(2007){Neto}, {Gao}, {Bett}, {Cole}, {Navarro},
  {Frenk}, {White}, {Springel}, \& {Jenkins}}]{neto.etal.07}
{Neto}, A.~F., {Gao}, L., {Bett}, P., {Cole}, S., {Navarro}, J.~F., {Frenk},
  C.~S., {White}, S.~D.~M., {Springel}, V., \& {Jenkins}, A. 2007, \mnras, 381,
  1450

\bibitem[{{Newman} {et~al.}(2013){Newman}, {Treu}, {Ellis}, \&
  {Sand}}]{newman.etal.13}
{Newman}, A.~B., {Treu}, T., {Ellis}, R.~S., \& {Sand}, D.~J. 2013, \apj, 765,
  25

\bibitem[{{Nord} {et~al.}(2008){Nord}, {Stanek}, {Rasia}, \&
  {Evrard}}]{nord.etal.08}
{Nord}, B., {Stanek}, R., {Rasia}, E., \& {Evrard}, A.~E. 2008, \mnras, 383,
  L10

\bibitem[{{Ogiya} \& {Mori}(2011)}]{ogiya.etal.11}
{Ogiya}, G. \& {Mori}, M. 2011, \apjl, 736, L2

\bibitem[{{Padovani} \& {Matteucci}(1993)}]{padovani&matteucci93}
{Padovani}, P. \& {Matteucci}, F. 1993, \apj, 416, 26

\bibitem[{{Pedrosa} {et~al.}(2009){Pedrosa}, {Tissera}, \&
  {Scannapieco}}]{pedrosa.etal.09}
{Pedrosa}, S., {Tissera}, P.~B., \& {Scannapieco}, C. 2009, \mnras, 395, L57

\bibitem[{{Piffaretti} \& {Valdarnini}(2008)}]{piffaretti&valdarnini}
{Piffaretti}, R. \& {Valdarnini}, R. 2008, \aap, 491, 71

\bibitem[{{Pointecouteau} {et~al.}(2005){Pointecouteau}, {Arnaud}, \&
  {Pratt}}]{pointecouteau.etal.05}
{Pointecouteau}, E., {Arnaud}, M., \& {Pratt}, G.~W. 2005, \aap, 435, 1

\bibitem[{{Power} {et~al.}(2003){Power}, {Navarro}, {Jenkins}, {Frenk},
  {White}, {Springel}, {Stadel}, \& {Quinn}}]{power.etal.03}
{Power}, C., {Navarro}, J.~F., {Jenkins}, A., {Frenk}, C.~S., {White},
  S.~D.~M., {Springel}, V., {Stadel}, J., \& {Quinn}, T. 2003, \mnras, 338, 14

\bibitem[{{Pratt} \& {Arnaud}(2005)}]{pratt&arnaud}
{Pratt}, G.~W. \& {Arnaud}, M. 2005, \aap, 429, 791

\bibitem[{{Pratt} {et~al.}(2009){Pratt}, {Croston}, {Arnaud}, \&
  {B{\"o}hringer}}]{pratt.etal.09}
{Pratt}, G.~W., {Croston}, J.~H., {Arnaud}, M., \& {B{\"o}hringer}, H. 2009,
  \aap, 498, 361

\bibitem[{{Ragone-Figueroa} {et~al.}(2012){Ragone-Figueroa}, {Granato}, \&
  {Abadi}}]{ragone.etal.12}
{Ragone-Figueroa}, C., {Granato}, G.~L., \& {Abadi}, M.~G. 2012, \mnras, 423,
  3243

\bibitem[{{Rasia} {et~al.}(2006){Rasia}, {Ettori}, {Moscardini}, {Mazzotta},
  {Borgani}, {Dolag}, {Tormen}, {Cheng}, \& {Diaferio}}]{rasia.etal.06}
{Rasia}, E., {Ettori}, S., {Moscardini}, L., {Mazzotta}, P., {Borgani}, S.,
  {Dolag}, K., {Tormen}, G., {Cheng}, L.~M., \& {Diaferio}, A. 2006, \mnras,
  369, 2013

\bibitem[{{Rasia} {et~al.}(2008){Rasia}, {Mazzotta}, {Bourdin}, {Borgani},
  {Tornatore}, {Ettori}, {Dolag}, \& {Moscardini}}]{rasia.etal.08}
{Rasia}, E., {Mazzotta}, P., {Bourdin}, H., {Borgani}, S., {Tornatore}, L.,
  {Ettori}, S., {Dolag}, K., \& {Moscardini}, L. 2008, \apj, 674, 728

\bibitem[{{Rasia} {et~al.}(2013){Rasia}, {Meneghetti}, \&
  {Ettori}}]{rasia.etal.12b}
{Rasia}, E., {Meneghetti}, M., \& {Ettori}, S. 2013, The Astronomical Review,
  8, 010000

\bibitem[{{Rasia} {et~al.}(2012){Rasia}, {Meneghetti}, {Martino}, {Borgani},
  {Bonafede}, {Dolag}, {Ettori}, {Fabjan}, {Giocoli}, {Mazzotta}, {Merten},
  {Radovich}, \& {Tornatore}}]{rasia.etal.12}
{Rasia}, E., {Meneghetti}, M., {Martino}, R., {Borgani}, S., {Bonafede}, A.,
  {Dolag}, K., {Ettori}, S., {Fabjan}, D., {Giocoli}, C., {Mazzotta}, P.,
  {Merten}, J., {Radovich}, M., \& {Tornatore}, L. 2012, New Journal of
  Physics, 14, 055018

\bibitem[{{Rasia} {et~al.}(2004){Rasia}, {Tormen}, \&
  {Moscardini}}]{rasia.etal.04}
{Rasia}, E., {Tormen}, G., \& {Moscardini}, L. 2004, \mnras, 351, 237

\bibitem[{{Rudd} {et~al.}(2008){Rudd}, {Zentner}, \& {Kravtsov}}]{rudd.etal.08}
{Rudd}, D.~H., {Zentner}, A.~R., \& {Kravtsov}, A.~V. 2008, \apj, 672, 19

\bibitem[{{Ryden} \& {Gunn}(1987)}]{ryden&gunn}
{Ryden}, B.~S. \& {Gunn}, J.~E. 1987, \apj, 318, 15

\bibitem[{{Schmidt} \& {Allen}(2007)}]{schmidt_allen}
{Schmidt}, R.~W. \& {Allen}, S.~W. 2007, \mnras, 379, 209

\bibitem[{{Sijacki} {et~al.}(2007){Sijacki}, {Springel}, {Di Matteo}, \&
  {Hernquist}}]{sijacki.etal.07}
{Sijacki}, D., {Springel}, V., {Di Matteo}, T., \& {Hernquist}, L. 2007,
  \mnras, 380, 877

\bibitem[{Springel(2005)}]{sp05.1}
Springel, V. 2005, \mnras, 364, 1105

\bibitem[{{Springel} {et~al.}(2005){Springel}, {Di Matteo}, \&
  {Hernquist}}]{springel.etal.05}
{Springel}, V., {Di Matteo}, T., \& {Hernquist}, L. 2005, \mnras, 361, 776

\bibitem[{Springel \& Hernquist(2003)}]{springel&hernquist03}
Springel, V. \& Hernquist, L. 2003, \mnras, 339, 289

\bibitem[{{Stanek} {et~al.}(2010){Stanek}, {Rasia}, {Evrard}, {Pearce}, \&
  {Gazzola}}]{stanek.etal.10}
{Stanek}, R., {Rasia}, E., {Evrard}, A.~E., {Pearce}, F., \& {Gazzola}, L.
  2010, \apj, 715, 1508

\bibitem[{{Stanek} {et~al.}(2009){Stanek}, {Rudd}, \&
  {Evrard}}]{stanek.etal.09}
{Stanek}, R., {Rudd}, D., \& {Evrard}, A.~E. 2009, \mnras, 394, L11

\bibitem[{{Teyssier} {et~al.}(2013){Teyssier}, {Pontzen}, {Dubois}, \&
  {Read}}]{teyssier.etal.12}
{Teyssier}, R., {Pontzen}, A., {Dubois}, Y., \& {Read}, J.~I. 2013, \mnras, 493

\bibitem[{{Tissera} {et~al.}(2010){Tissera}, {White}, {Pedrosa}, \&
  {Scannapieco}}]{tissera.etal.10}
{Tissera}, P.~B., {White}, S.~D.~M., {Pedrosa}, S., \& {Scannapieco}, C. 2010,
  \mnras, 406, 922

\bibitem[{Tormen {et~al.}(1997)Tormen, Bouchet, \& White}]{tormen.etal.97}
Tormen, G., Bouchet, F., \& White, S. D.~M. 1997, \mnras, 286, 865

\bibitem[{{Tornatore} {et~al.}(2007){Tornatore}, {Borgani}, {Dolag}, \&
  {Matteucci}}]{tornatore.etal.07}
{Tornatore}, L., {Borgani}, S., {Dolag}, K., \& {Matteucci}, F. 2007, \mnras,
  382, 1050

\bibitem[{{van Daalen} {et~al.}(2011){van Daalen}, {Schaye}, {Booth}, \& {Dalla
  Vecchia}}]{vandaalen.etal.11}
{van Daalen}, M.~P., {Schaye}, J., {Booth}, C.~M., \& {Dalla Vecchia}, C. 2011,
  \mnras, 415, 3649

\bibitem[{{Vikhlinin} {et~al.}(2009){Vikhlinin}, {Burenin}, {Ebeling},
  {Forman}, {Hornstrup}, {Jones}, {Kravtsov}, {Murray}, {Nagai}, {Quintana}, \&
  {Voevodkin}}]{vikh.etal.09}
{Vikhlinin}, A., {Burenin}, R.~A., {Ebeling}, H., {Forman}, W.~R., {Hornstrup},
  A., {Jones}, C., {Kravtsov}, A.~V., {Murray}, S.~S., {Nagai}, D., {Quintana},
  H., \& {Voevodkin}, A. 2009, \apj, 692, 1033

\bibitem[{{Vikhlinin} {et~al.}(2006){Vikhlinin}, {Kravtsov}, {Forman}, {Jones},
  {Markevitch}, {Murray}, \& {Van Speybroeck}}]{vikh.etal.06}
{Vikhlinin}, A., {Kravtsov}, A., {Forman}, W., {Jones}, C., {Markevitch}, M.,
  {Murray}, S.~S., \& {Van Speybroeck}, L. 2006, \apj, 640, 691

\bibitem[{{Voit}(2005)}]{voit05}
{Voit}, G.~M. 2005, Reviews of Modern Physics, 77, 207

\bibitem[{{Walker} {et~al.}(2013){Walker}, {Fabian}, {Sanders}, {Simionescu},
  \& {Tawara}}]{walker.etal.13}
{Walker}, S.~A., {Fabian}, A.~C., {Sanders}, J.~S., {Simionescu}, A., \&
  {Tawara}, Y. 2013, \mnras, 432, 554

\bibitem[{{Wiersma} {et~al.}(2009){Wiersma}, {Schaye}, \&
  {Smith}}]{wiersma.etal.09}
{Wiersma}, R.~P.~C., {Schaye}, J., \& {Smith}, B.~D. 2009, \mnras, 393, 99

\bibitem[{{Zemp} {et~al.}(2012){Zemp}, {Gnedin}, {Gnedin}, \&
  {Kravtsov}}]{zhemp.etal.12}
{Zemp}, M., {Gnedin}, O.~Y., {Gnedin}, N.~Y., \& {Kravtsov}, A.~V. 2012, \apj,
  748, 54

\bibitem[{{Zhao} {et~al.}(2009){Zhao}, {Jing}, {Mo}, \&
  {B{\"o}rner}}]{zhao.etal.09}
{Zhao}, D.~H., {Jing}, Y.~P., {Mo}, H.~J., \& {B{\"o}rner}, G. 2009, \apj, 707,
  354

\bibitem[{{Zhu} \& {Pan}(2012)}]{zhu.etal.12}
{Zhu}, X.-J. \& {Pan}, J. 2012, Research in Astronomy and Astrophysics, 12, 2

\end{thebibliography}

\end{document}